\documentclass[12pt,english,aip,jcp]{revtex4}
\usepackage{graphicx}
\newcommand{\I}{\mathbf{1}}
\newcommand{\bra}[1]{\langle #1|}
\newcommand{\ket}[1]{|#1\rangle}

\usepackage{booktabs}
\usepackage{multirow}
\usepackage{lscape}

\begin{document}

\title{The Bravyi-Kitaev transformation for quantum computation of electronic structure }

\author{Jacob T. Seeley, Martin J. Richard, Peter J. Love \\
{\it Haverford College}\\
{\it Department of Physics}\\
{\it 370 Lancaster Ave}\\
{\it Haverford, PA 19041}\\
}
\date{\today}

\begin{abstract}
Quantum simulation is an important application of future quantum computers with applications in quantum chemistry, condensed matter, and beyond. Quantum simulation of fermionic systems presents a specific challenge. The Jordan-Wigner transformation allows for representation of a fermionic operator by $O(n)$ qubit operations. Here we develop an alternative method of simulating fermions with qubits, first proposed by Bravyi and Kitaev [S. B. Bravyi, A.Yu. Kitaev, {\it Annals of Physics} 298, 210-226 (2002)],  that reduces the simulation cost to $O(\log n)$ qubit operations for one 
fermionic operation. We apply this new Bravyi-Kitaev transformation to the task of simulating quantum chemical Hamiltonians,
and give a detailed example for the simplest possible case of molecular hydrogen in a minimal basis. We show that the quantum circuit for 
simulating a single Trotter time-step of the Bravyi-Kitaev derived Hamiltonian for H$_2$ requires fewer gate applications than the equivalent circuit derived from the Jordan-Wigner transformation. Since the scaling of the Bravyi-Kitaev method is asymptotically better than the Jordan-Wigner method, this result for molecular hydrogen
in a minimal basis demonstrates the superior efficiency of the Bravyi-Kitaev method for all quantum computations of electronic structure.
\end{abstract}

\maketitle

\section{Introduction}\label{intro}


In his seminal article that anticipated the field of quantum information, Feynman argued that simulating quantum systems on classical computers takes an amount of time that scales exponentially with the size of the system, while the cost of quantum simulations can scale in polynomial time with system size \cite{Feynman}. This possibility may offer a path forward for computational chemistry \cite{Lloyd, Alan}.  A quantum simulation algorithm for quantum chemical Hamiltonians enables the efficient calculation of properties such as energy spectra \cite{Alan}, reaction rates \cite{Dan, Ivan}, correlation functions \cite{Ortiz}, and molecular properties \cite{Ivan2} for molecules larger than those that are currently accessible through classical calculations.

Quantum simulation of electronic structure requires a representation of fermions by systems of qubits. Significant progress has been made on efficient quantum simulation of fermions. In 1997, Abrams and Lloyd proposed a simulation scheme for fermions hopping on a lattice~\cite{AL}.  In 2002, Somma {\it et al}. used the Jordan-Wigner to generalize the simulation scheme proposed by Abrams and Lloyd \cite{Somma, JW}. The Jordan-Wigner transformation has since been used to outline a scalable quantum algorithm for the simulation of molecular electron dynamics, and to design an explicit quantum circuit for simulating a Trotter time-step of the molecular electronic Hamiltonian for H$_2$ in a minimal basis \cite{Alan, James}. Further refinements of the Jordan-Wigner construction were made by Verstrate and Cirac \cite{VC} and by Bravyi and Kitaev \cite{BK}. From the point of view of fundamental physics, such constructions can be regarded as giving a negative answer to the question of whether fundamental fermi fields are required to explain observed fermionic degrees of freedom~\cite{Ball}. Practically speaking, such constructions show that quantum computation of electronic structure does not suffer from an analog of the sign problem; that is, fermion antisymmetry represents no significant obstacle to efficient algorithms. 

Theoretical progress in quantum simulation has been accompanied by experimental successes. In 2010, Lanyon {\it et al}.\ calculated the energy spectrum of a hydrogen molecule using an optical quantum computer~\cite{Lanyon}. For a review of photonic quantum simulators, see \cite{AlanWalther}. Du {\it et al}.\ repeated this result to higher precision with NMR shortly thereafter \cite{Du}. Digital quantum simulations of the kind considered in the present paper have been implemented in ion traps using up to 100 gates and 6 qubits~\cite{Blatt2}. The progress of trapped ion quantum simulation is detailed in \cite{Blatt}. 

Quantum computation of electronic structure has been the subject of simulation studies~\cite{Alan,Veis1} and has been extended to cover relativistic systems~\cite{Veis2}. The history of calculations in quantum chemistry provides a useful sequence of problems reaching from calculations that can be performed on experimental quantum computers today to calculations at the present research frontier~\cite{Love}. Despite these promising results, the scaling of the number of gates required by the algorithm outlined in \cite{Alan, James} remains challenging. It is a subject of active research to find improvements to the (polynomial) scaling of the cost of the algorithm described in~\cite{Alan, James}. Several improvements are described in~\cite{Jones}, and the techniques of that work could be combined with those of the present paper to further reduce the resource requirements. 

\begin{figure}[ht]
 \centerline{\includegraphics[scale = .45]{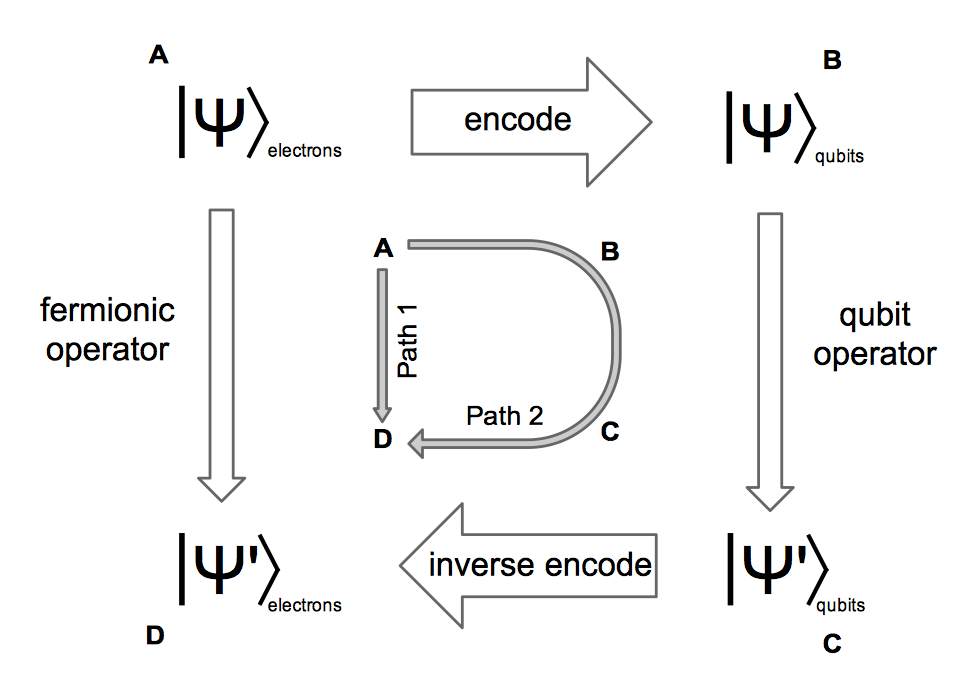}}
 \caption[Criterion for a successful simulation scheme]{ A simulation scheme first encodes fermionic states in qubits, then acts with the qubit operator representing the fermionic operator (obtained by the associated transformation), then inverts the encoding to obtain the resultant fermionic state. The criterion for a successful simulation scheme is that this procedure reproduces the action of the fermionic operator, i.e. that Path 1 is equivalent to Path 2, for all basis states --- in other words, that this diagram commutes.\label{fig1}}
\end{figure}

A fermionic simulation scheme can be broken into two pieces: first, to map occupation number basis vectors to states of qubits; and second, to represent the fermionic creation and annihilation operators in terms of operations on qubits in a way that preserves the fermionic anti-commutation relations, as illustrated in Figure~\ref{fig1}. Previous simulation algorithms have used a straightforward mapping of fermionic occupation number basis states to qubit states that was originally defined by Zanardi in the context of entanglement \cite{Zanardi,Somma,Alan}. The Jordan-Wigner transformation is then used to write the electronic Hamiltonian as a sum over products of Pauli spin operators acting on the qubits of the quantum computer. Subsequently the Hamiltonian terms $h_k$, where $\hat{H} = \sum_{k} h_k$, are converted into the unitary gates that are the corresponding time evolution operators. Even though the $h_k$ do not necessarily commute, their sequential execution on a quantum computer can be made to approximate the unitary propagator $e^{-i \hat{H} t}$ through a Trotter decomposition \cite{Trotter, Suzuki, QA,MikeIke}. Finally, the iterative phase estimation algorithm (IPEA) is used to approximate the eigenvalue of an input eigenstate \cite{Alan, James, MikeIke}. 

In this paper we treat the Trotterization process and IPEA as standard procedures. We develop the Bravyi-Kitaev basis and Bravyi-Kitaev transformation, both named after the authors who first proposed such a scheme \cite{BK}, which provide a more efficient mapping between electronic Hamiltonians and qubit Hamiltonians.  While the occupation number basis and the Jordan-Wigner transformation allow for the representation of a single fermionic creation or annihilation operator by $O(n)$ qubit operations, the Bravyi-Kitaev basis and transformation require only $O(\log n)$ qubit operations to represent one fermionic operator. It is worth noting that Bravyi and Kitaev were concerned with exploring the power of fermions as the basic hardware units of a quantum computer, rather than with the simulation of fermions by qubits \cite{BK}. However, understanding how the structure of fermionic systems can be employed to process information helps us understand how standard quantum information procedures can be used to simulate the structure of fermionic systems. We work out a detailed application of the Bravyi-Kitaev transformation to the operators that appear in quantum chemical Hamiltonians, providing a new way of mapping electronic Hamiltonians to qubit Hamiltonians. We also give explicit Pauli decompositions of the qubit operators derived from this new transformation for the quantum chemical Hamiltonian for H$_2$ in a minimal basis. We show that the quantum circuit for simulating a single first-order Trotter time-step of the Bravyi-Kitaev minimal basis molecular hydrogen Hamiltonian requires 30 single-qubit gates and 44 CNOT gates, as compared to 46 single-qubit gates and 36 CNOT gates for the Jordan-Wigner Hamiltonian derived in \cite{James}. Finally, we show that a chemical-precision estimate of the ground state eigenvalue of the Bravyi-Kitaev Hamiltonian can be obtained in 3 first-order Trotter steps, with a total cost of 222 gates, while the Jordan-Wigner Hamiltonian requires 4 first-order Trotter steps for a total of 328 gates. Since the Bravyi-Kitaev transformation is known to be asymptotically more efficient, this result for the simplest possible case of molecular hydrogen in a minimal basis demonstrates the superior efficiency of the Bravyi-Kitaev method for all molecular quantum simulations.

In Section~\ref{back} we will review basic quantum chemistry in second quantized form as well as the Jordan Wigner transformation. In Section~\ref{alt} we discuss alternatives to the occupation number basis, including the Bravyi-Kitaev basis, which we go on to describe in detail in Section~\ref{sets}. In Section~\ref{bravkit} we present the Bravyi-Kitaev transformation, which allows us to represent creation and annihilation operators in the Bravyi-Kitaev basis. In Section~\ref{pauli} we compute the products of these operators that occur in electronic structure Hamiltonians. In Section~\ref{molham} we compute the molecular electronic structure Hamiltonian of H$_2$ in a minimal basis using the Bravyi-Kitaev basis and transformation. In Section~\ref{trott} we make an explicit comparison between the Bravyi-Kitaev transformation and the Jordan Wigner transformation by simulating the Trotterization procedure. We close the paper with some conclusions about the utility of the Bravyi-Kitaev transformation.

\section{Background}\label{back}

\subsection{Fermionic systems and second quantization}

We may describe fermionic systems using the formalism of second quantization, in which $n$ single-particle states can be either empty or occupied by a spinless fermionic particle. In the context of quantum chemistry these $n$ states represent spin orbitals, ideally one-electron energy eigenfunctions and often molecular orbitals found by the Hartree-Fock method~\cite{Mcweeny,Szabo}. We consider a subspace of the full Fock space which is spanned by $2^n$ electronic basis states $\ket{f_{n-1}\ \ldots\ f_{0}}$, where $f_j \in \{0,1\}$ is the occupation number of orbital $j$ (restricted to these values due to the Pauli exclusion principle). This is called the {\em occupation number} basis.

Any interaction of a fermionic system can be expressed in terms of products of the creation and annihilation operators $a^\dag_j$ and $a_j$, for $j \in \{0,\ldots,n\!-\!1\}$. Due to the exchange anti-symmetry of fermions, the action of $a^\dag_j$ or $a_j$ introduces a phase to the electronic basis state that depends on the occupancy of all orbitals with index less than $j$ in the occupation number representation. (One can choose instead to define these operators so that it is the occupation of orbitals with 
index greater than $j$ that determines the phase --- the ordering of orbitals is arbitrary.) These operators act on occupation number basis vectors as follows:

\begin{eqnarray}
a^\dag_j\, \ket{f_{n-1}\  \ldots\  f_{j+1}\  0\  f_{j-1}\ \ldots\ f_{0}}
& \,=\, \bigl(-1\bigr)^{\sum_{s=0}^{j-1}f_s}\,
\ket{f_{n-1}\ \ldots\ f_{j+1}\ 1\ f_{j-1}\ \ldots\ f_{0}}; \\
a^\dag_j\, \ket{f_{n-1}\ \ldots\ f_{j+1}\ 1\ f_{j-1}\ \ldots\ f_{0}}
& \,=\, 0;\\
a_j\, \ket{f_{n-1}\  \ldots\  f_{j+1}\  1\  f_{j-1}\ \ldots\ f_{0}}
& \,=\, \bigl(-1\bigr)^{\sum_{s=0}^{j-1}f_s}\,
\ket{f_{n-1}\ \ldots\ f_{j+1}\ 0\ f_{j-1}\ \ldots\ f_{0}}; \\
a_j\, \ket{f_{n-1}\ \ldots\ f_{j+1}\ 0\ f_{j-1}\ \ldots\ f_{0}}
& \,=\, 0.
\end{eqnarray}
The canonical fermionic anti-commutation relations enforce the exchange anti-symmetry:
\begin{equation} [a_j,a_k]_+=0, \qquad [a^\dag_j,a^\dag_k]_+=0, \qquad [a_j,a^\dag_k]_+=\delta_{jk}\I, \end{equation}
where the anti-commutator of operators $A$ and $B$ is defined by $[A,B]_+\equiv AB+BA$.

The molecular electronic Hamiltonian of interest in the electronic structure problem is:
\begin{equation}\label{molhameq}
\hat{H}=\sum_{i,j}h_{ij}\ a^\dag_i a_j +\frac{1}{2}\sum_{i,j,k,l} h_{ijkl}\ a^\dag_i a^\dag_j a_k a_l.
\end{equation}
The coefficients $h_{ij}$ and $h_{ijkl}$ are one- and two-electron overlap integrals, which can be precomputed classically and input to the quantum simulation as parameters~\cite{Alan,James, Mcweeny}. 

As an application of the techniques presented in this paper (Section~\ref{molham}), we treat molecular hydrogen in a minimal basis. Thus, we construct two spatial molecular orbitals by taking linear combinations of the localized atomic spatial wavefunctions: $\psi_g = \psi_{H1} + \psi_{H2}$ and $\psi_u = \psi_{H1} - \psi_{H2}$. Here the subscripts {\it g} and {\it u} stand for the German words {\it gerade} and {\it ungerade} --- even and odd. In general one must take a Slater determinant to determine the correctly anti-symmetrized wavefunctions of the fermionic system, but in this case we can guess them by inspection. The form of the spatial wavefunctions is determined by the choice of basis set. STO-3G is a commonly used Gaussian basis set --- for further details see \cite{Mcweeny,Szabo}.

Molecular spin orbitals are formed by taking the product of these two molecular spatial orbitals with one of two orthogonal spin functions, $\ket{\alpha}$ and $\ket{\beta}$. Thus, the four molecular spin orbitals in our model of the hydrogen molecule (which correspond to the operators $a_j^{(\dag)}$) are:
\begin{equation}
\ket{\chi_0} = \ket{\psi_g} \ket{\alpha}, \qquad \ket{\chi_1} = \ket{\psi_g} \ket{\beta}, \qquad \ket{\chi_2} = \ket{\psi_u} \ket{\alpha}, \qquad \ket{\chi_3} = \ket{\psi_u} \ket{\beta}.
\end{equation}
In the next section we will review the occupation number basis and the Jordan-Wigner transformation, which together have been established as
a standard method for mapping fermionic systems to quantum computers~\cite{Alan,Somma,James,Lanyon}.

\subsection{The Jordan-Wigner transformation}
The form of electronic occupation number basis vectors suggests the following identification between 
electronic basis states on the left and states of our quantum computer~\cite{Zanardi}:
\begin{equation} \ket{f_{n-1}\ \ldots\ f_1\ f_{0}} \rightarrow
\ket{q_{n-1}} \cdots \otimes \ket{q_1}\otimes \ket{q_0}, \qquad\ f_j = q_j \in \{0,1\}.
\end{equation}
That is, we let the state of each qubit $\ket{q_j}$ store $f_j$, the occupation number of orbital $j$. We refer to this method of encoding fermionic states as the occupation number basis for qubits. The next step is to map fermionic creation and annihilation operators onto operators on qubits.

We can form one-qubit creation and annihilation operators, $\hat{Q}^{+}$ and $\hat{Q}^{-}$, that act on qubits of our quantum computer as follows:
\begin{eqnarray}
& \hat{Q}^{+}\ket{0}=\ket{1},  \qquad  \hat{Q}^{+}\ket{1}=0,  \qquad \hat{Q}^{-} \ket{1}=\ket{0},  \qquad  \hat{Q}^{-} \ket{0}=0.
\end{eqnarray}
We could proceed by following the standard recipe for turning $p$-qubit quantum gates into operators acting on an $n$-qubit 
quantum computer ($n \geq p$) by taking the tensor product of the gates acting on the target qubits with the identity
acting on the other ($n-p$) qubits. However, it is easy to show that the qubit creation and annihilation operators formed in this way
do not obey the fermionic anti-commutation relations.

Expressing the qubit creation and annihilation operators in terms of Pauli matrices suggests a way forward:
\begin{equation}
\hat{Q}^{+} = \ket{1}\bra{0} = \frac{1}{2}(\sigma^x - i\sigma^y), \qquad
\hat{Q}^{-} = \ket{0}\bra{1} = \frac{1}{2}(\sigma^x + i\sigma^y).
\end{equation}
The mutual anti-commutation of the three Pauli matrices allows us to recognize that $\hat{Q}^{\pm}$ anti-commutes with $\sigma^z$. Thus if we represent the action of $a^\dag_j$ or $a_j$ by acting with $\hat{Q}^{\pm}_{j}$ and with $\sigma^z$ on all qubits with index less than $j$, our qubit operators will obey the fermionic anti-commutation relations. Put differently, the states of our quantum computer will acquire the same phases under the action of our qubit operator as do the electronic basis states under the action of the corresponding creation or annihilation operator. The effect of the string of $\sigma^z$ gates is to introduce the required phase change of $-1$ if the parity of the set of qubits with index less than $j$ is 1 (odd), and to do nothing if the parity is 0 (even), where the parity of a set of qubits is just the sum ($\bmod ~2$) of the numbers that represent the states they are in.

We can then completely represent the fermionic creation and annihilation operators in terms of basic qubit gates as follows:
\begin{equation}
    a_j^\dagger\equiv{\I}^{\otimes n-j-1}\otimes \hat{Q}^{+} \otimes [{\sigma^z}^{\otimes j}], \qquad a_j\equiv{\I}^{\otimes n-j-1}\otimes \hat{Q}^{-} \otimes [{\sigma^{z}}^{\otimes j}].
\end{equation}
A more compact notation, of which we will make extensive use throughout this paper, is:
\begin{eqnarray}
&a_j^\dagger \equiv \hat{Q}^{+}_j \otimes Z^{\rightarrow}_{j-1} = \frac{1}{2} (X_j \otimes Z^{\rightarrow}_{j-1} - i Y_j \otimes Z^{\rightarrow}_{j-1}); \\
&a_j \equiv \hat{Q}^{-}_j \otimes  Z^{\rightarrow}_{j-1} = \frac{1}{2} (X_j \otimes Z^{\rightarrow}_{j-1} + i Y_j \otimes Z^{\rightarrow}_{j-1}),
\end{eqnarray}
where:
\begin{equation} 
Z^{\rightarrow}_{i} \equiv \sigma^z_i \otimes \sigma^z_{i-1} \otimes \cdots \sigma^z_1 \otimes \sigma^z_0,
\end{equation}
and where it is assumed that any qubit not explicitly operated on is acted on by the identity. The operator $Z^{\rightarrow}_i$ is a ``parity operator" with eigenvalues $\pm 1$, corresponding to eigenstates for which the subset of bits with index less than or equal to $i$ has even or odd parity, respectively.

The above correspondence, a mapping of interacting fermions to spins, is the Jordan-Wigner transformation \cite{Alan,JW,James, Aeppli}. Jordan and Wigner introduced this transformation in 1928 in the context of 1D lattice models, but it has since been applied to quantum simulation of fermions \cite{Alan,Somma, JW, James}. The problem with this method is that as a consequence of the non-locality of the parity operator $Z^{\rightarrow}_i$, the number of extra qubit operations required to simulate a single fermionic operator scales as $O(n)$. In the next section we consider two alternatives to the occupation number basis that were suggested by Bravyi and Kitaev \cite{BK}.

\section{Alternatives to the occupation number basis}\label{alt}
\subsection{The parity basis}
The extra qubit operations required to simulate one fermionic operator when using the Jordan-Wigner method result from operating with $\sigma^z$ on all qubits with index less than $j$. This task could be accomplished by a single application of $\sigma^z$ if instead of using qubit $j$ to store $f_j$, we used qubit $j$ to store the {\em parity} of all occupied orbitals up to orbital $j$ \cite{BK}. That is, we could let qubit $j$ store $p_j=\sum_{s=0}^{j}f_s$. (Throughout this paper, all sums of binary variables are taken $\bmod ~2$). We follow~\cite{BK} and call this encoding of fermionic states in qubit states the {\em{parity}} basis. 

It is useful to define the transformations between bases we will consider in terms of maps between bit strings. For all the transformations we consider, which involve only sums of bits $\bmod ~2$, it is possible to represent their action by matrices acting on the vector of bit values corresponding to a given logical basis state. For example, the occupation number basis state $\ket{f_7 \ldots f_1f_0}$ is equivalent to the following vector:

\begin{equation}
(f_7, \dots, f_1, f_0)^T \\
\end{equation}

In terms of these vectors the map to the parity basis is given by:
\begin{equation}\label{threeone}
p_i = \sum_j [ \pi_n]_{ij}\ f_j,
\end{equation}
where $n$ is the number of orbitals. $\pi_n$ is the $(n\times n)$ matrix defined below. Note that we index the matrix $\pi_n$ from the lower right corner, for consistency with our orbital numbering scheme. 
\begin{equation}
  [\pi_n]_{ij} = \left\{ 
  \begin{array}{l l}
    1 & \quad i < j \\
    0 & \quad i \geq j\\
  \end{array} \right. ,
  \qquad
  {\rm so~that}
  \qquad
  \pi_n = \left(
 \begin{array}{cccc}
  1 & 1 & \cdots & 1 \\
  0 & 1 & \cdots & 1 \\
  \vdots  & \vdots  & \ddots & \vdots  \\
  0 & 0 & \cdots & 1
 \end{array}
 \right)
\end{equation}
For example, to change the occupation number basis state $\ket{1 0 1 0 0 1 1 1}$ into its corresponding parity basis state $\ket{1 0 0 1 1 1 0 1}$, we act with the matrix $\pi_8$ on the appropriate bit string:
\begin{equation}
\begin{bordermatrix}{~ & f_7 & f_6 & f_5 & f_4 & f_3 & f_2 & f_1 & f_0   \cr
                  p_7 & 1 & 1 & 1 & 1 & 1 & 1 & 1 & 1 \cr
                  p_6 & 0 & 1 & 1 & 1 & 1 & 1 & 1 & 1 \cr
                  p_5 & 0 & 0 & 1 & 1 & 1 & 1 & 1 & 1 \cr
                  p_4 & 0 & 0 & 0 & 1 & 1 & 1 & 1 & 1 \cr
                  p_3 & 0 & 0 & 0 & 0 & 1 & 1 & 1 & 1 \cr
                  p_2 & 0 & 0 & 0 & 0 & 0 & 1 & 1 & 1 \cr
                  p_1 & 0 & 0 & 0 & 0 & 0 & 0 & 1 & 1 \cr
                  p_0 & 0 & 0 & 0 & 0 & 0 & 0 & 0 & 1 \cr}
 \end{bordermatrix}
\left(\begin{matrix}
  {1 \cr
  0 \cr
  1 \cr
  0 \cr
  0 \cr
  1 \cr
  1 \cr
  1\cr}
 \end{matrix}
 \right)
 =
 \begin{pmatrix}
 { 1 \cr
  0 \cr
  0 \cr
  1 \cr
  1 \cr
  1 \cr
  0 \cr
  1\cr}
 \end{pmatrix}
\end{equation}
With this understanding of the parity basis transformation, we can now derive the transformation that maps fermionic operators into operators in the parity basis. Since the parity of the set of orbitals with index less than $j$ is what determines whether the action of $a^{(\dag)}_j$ introduces a phase of $-1$, operating with $\sigma^z$ on qubit $(j-1)$ alone will introduce the necessary phase to the corresponding qubit state in the parity basis.

However, unlike the Jordan-Wigner transformation, we cannot represent the creation or annihilation of a particle in orbital $j$ by simply operating
with $\hat{Q}^{\pm}$ on qubit $j$, because in the parity basis qubit $j$ does not store the occupation of orbital $j$, but the parity of all
orbitals with index less than or equal to $j$. Thus whether we need to act with $\hat{Q}^{+}$ or $\hat{Q}^{-}$ on qubit $j$
depends on qubit $(j-1)$. If qubit $(j-1)$ is in the state $\ket{0}$, then qubit $j$ will accurately reflect the
occupation of orbital $j$, and simulating $a^\dag_j$ will require acting on qubit $j$ with $\hat{Q}^{+}$, as before. But if qubit $(j-1)$ is in the state $\ket{1}$,
then qubit $j$ will have inverted parity compared to the occupation of orbital $j$, and we will instead need to act with $\hat{Q}^{-}$ on qubit $j$ to simulate $a^\dag_j$ (and {\em vice versa} for the annihilation operator).

The operator equivalent to $\hat{Q}^{\pm}$ in the parity basis is therefore a two-qubit operator acting on qubits $j$ and $j-1$:
\begin{equation}
\hat{\mathcal{P}}^{\pm}_j \equiv \hat{Q}^{\pm}_j \otimes \ket{0}\bra{0}_{j-1} - \hat{Q}^{\mp}_j \otimes \ket{1}\bra{1}_{j-1} = \frac{1}{2}(X_j \otimes Z_{j-1} \mp i Y_j).
\end{equation}
Additionally, creating or annihilating a particle in orbital $j$ changes the parity data that must be stored by
all qubits with index greater than $j$. Thus we must update the cumulative sums $p_k$ for $k > j$ by applying
$\sigma^x$ to all qubits $\ket{p_k}$, $k > j$ \cite{BK}. The representations of the creation and annihilation operators in the parity basis are then:

\begin{eqnarray}
&a_j^\dagger \equiv X^{\leftarrow}_{j+1} \otimes \hat{\mathcal{P}}^{+}_j = \frac{1}{2} (X^{\leftarrow}_{j+1} \otimes X_j \otimes Z_{j-1} - i X^{\leftarrow}_{j+1} \otimes Y_j); \\
&a_j \equiv X^{\leftarrow}_{j+1} \otimes \hat{\mathcal{P}}^{-}_j = \frac{1}{2} (X^{\leftarrow}_{j+1} \otimes X_j \otimes Z_{j-1} + i X^{\leftarrow}_{j+1} \otimes Y_j),
\end{eqnarray}
where:
\begin{equation}
X^{\leftarrow}_{i} \equiv \sigma^x_{n-1} \otimes \sigma^x_{n-2} \otimes \cdots \sigma^x_{i+1} \otimes \sigma^x_{i}.
\end{equation}
This is the equivalent of the Jordan-Wigner transformation for the parity basis. The operator $X^{\leftarrow}_{i}$ is the ``update operator", which updates all qubits that store a partial sum including orbital $(i-1)$ when the occupation number of that orbital changes. It is straightforward to verify that these mappings satisfy the fermionic anti-commutation relations. But to simulate fermionic operators in the parity basis, we have traded the trailing string of $\sigma^z$ gates required by the
Jordan-Wigner transformation for a leading string of $\sigma^x$ gates whose length also scales as $O(n)$, and we have
not improved on the efficiency of the Jordan-Wigner simulation procedure. In the next section, we explore a third possibility.

\subsection{The Bravyi-Kitaev basis}
Two kinds of information are required to simulate fermionic operators with qubits: the occupation of the target orbital, and the parity of the set of orbitals with index less than the target orbital. The previous two approaches are dual in the way that they store this information. With the occupation number basis and its associated Jordan-Wigner transformation, the occupation information is stored locally but the parity information is non-local, whereas in the parity basis method and its corresponding operator transformation, the parity information is stored locally but the occupation information is non-local. 

The Bravyi-Kitaev basis is a middle ground.
That is, it balances the locality of occupation and parity information for improved simulation efficiency. The general form of such a scheme
must be to use qubits $\ket{b_j}$ to store $partial$ sums $\sum_{s=k}^{l}f_s$ of occupation numbers according to some
algorithm. For ease of explanation, in the exposition that follows, when we write that a qubit ``stores a set of orbitals", what is meant is that the qubit
stores the parity of the set of occupation numbers corresponding to that set of orbitals.

Bravyi and Kitaev's encoding has an elegant binary grouping structure \cite{BK}. In this scheme, qubits store the parity of a set of $2^x$ orbitals, where $x \geq 0$. A qubit of index $j$ always stores orbital $j$. For even values of $j$, this is the only orbital that it stores, but for odd values of $j$, it also stores a certain set of adjacent orbitals with index less than $j$. Just as with the parity basis transformation, this encoding can be symbolized in a matrix $\beta_n$ that acts on bit string vectors corresponding to occupation number basis vectors of length $n$ to transform them to the corresponding Bravyi-Kitaev-encoded bit strings (again, all additions done ${\bmod \ 2}$). In terms of these vectors, the map from the occupation number basis to the Bravyi-Kitaev basis is:
\begin{equation}
b_i = \sum_j [ \beta_n]_{ij}\ f_j,
\end{equation}
where the matrix $\beta_n$ is given in Figure~\ref{figtwo} below.

\begin{figure}[!ht]
 \centerline{\includegraphics[scale = .5]{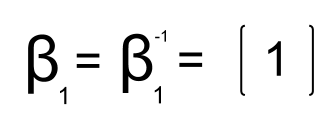}}
 \centerline{\includegraphics[scale = .5]{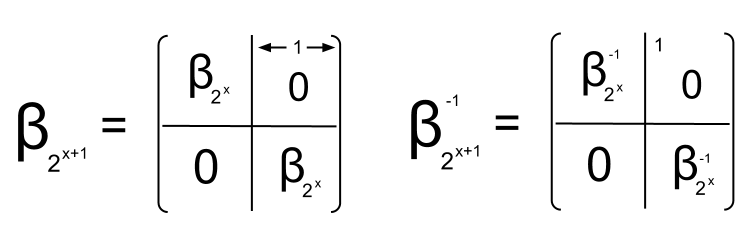}}
 \caption[Change of basis matrix (from occupation to Bravyi-Kitaev)]{\label{figtwo}The matrix $\beta_n$ that transforms occupation number basis vectors of length $n$ into the Bravyi-Kitaev basis. $\beta_1$ is a $(1 \times 1)$ matrix with a single entry of 1. Subsequent iterations of the matrix that act on occupation number basis vectors
of length $2^x$ are constructed by taking $\I \otimes \beta_{2^{x-1}}$ and then filling in the top row of the first quadrant of this matrix with 1's.
$\beta_n$ for $2^x < n < 2^{x+1}$ is just the $(n \times n)$ segment of $\beta_{2^{x+1}}$ that includes $b_0$ through
$b_{n-1}$. The recursion pattern for the inverse transformation matrix is also shown. An entry of 1 in row $b_i$, column $f_j$ means that $b_i$ is a partial sum including $f_j$.}

\end{figure}

For example, to change the occupation number basis state $\ket{1 0 1 0 0 1 1 1}$ into its corresponding Bravyi-Kitaev basis state $\ket{1 0 1 0 1 1 0 1}$, we act with the matrix $\beta_8$ on the appropriate bit string vector:
\begin{equation}
\begin{bordermatrix}{~ & f_7 & f_6 & f_5 & f_4 & f_3 & f_2 & f_1 & f_0   \cr
                  p_7 & 1 & 1 & 1 & 1 & 1 & 1 & 1 & 1 \cr
                  p_6 & 0 & 1 & 1 & 1 & 1 & 1 & 1 & 1 \cr
                  p_5 & 0 & 0 & 1 & 1 & 1 & 1 & 1 & 1 \cr
                  p_4 & 0 & 0 & 0 & 1 & 1 & 1 & 1 & 1 \cr
                  p_3 & 0 & 0 & 0 & 0 & 1 & 1 & 1 & 1 \cr
                  p_2 & 0 & 0 & 0 & 0 & 0 & 1 & 1 & 1 \cr
                  p_1 & 0 & 0 & 0 & 0 & 0 & 0 & 1 & 1 \cr
                  p_0 & 0 & 0 & 0 & 0 & 0 & 0 & 0 & 1 \cr}
 \end{bordermatrix}
\left(\begin{matrix}
  {1 \cr
  0 \cr
  1 \cr
  0 \cr
  0 \cr
  1 \cr
  1 \cr
  1\cr}
 \end{matrix}
 \right)
 =
 \left(\begin{matrix}
 { 1 \cr
  0 \cr
  1 \cr
  0 \cr
  1 \cr
  1 \cr
  0 \cr
  1\cr}
 \end{matrix}
 \right)
\end{equation}

This encoding strikes a balance between the occupation number basis and the parity basis
methods. The parity of occupied orbitals up to orbital $j$ is no longer stored in a single qubit, but the Bravyi-Kitaev
encoding stores the parity of orbitals with index less than $j$ in a few partial sums whose number scales as $O(\log j) \leq O(\log n)$ \cite{BK}.
Likewise, we no longer need to update {\em all} the qubits with index
greater than $j$, but only those that store partial sums which include occupation number $j$. Each
occupation number enters an additional partial sum only if the number of single particle states $n$ is doubled, and so
the overall cost of simulating a single fermionic operator with qubits scales as $O(\log n)$ \cite{BK}.

Given this encoding, we need to determine --- for an arbitrary index $j$ --- which qubits in the Bravyi-Kitaev basis store the parity of all orbitals with index less than $j$, which qubits store a partial sum including orbital $j$, and which qubits determine whether qubit $j$ has the same parity or inverted parity with respect to orbital $j$. These sets of indices will allow us to explicitly construct the fermionic creation and annihilation operators in the Bravyi-Kitaev basis. In the next section, we define these sets of qubit indices.

\section{Sets of qubits relevant to the Bravyi-Kitaev basis}\label{sets}
In this section we define the sets of qubits that are involved in the Bravyi-Kitaev transformation. These are the {\em parity set} (the qubits in the Bravyi-Kitaev basis that store the parity of all orbitals with index less than $j$), the {\em update set} (the qubits that store a partial sum including orbital $j$), and the {\em flip set} (the qubits that determine whether qubit $j$ has the same parity as orbital $j$).

\subsection{The parity set}
For an arbitrary index $j$, we would like to know which set of qubits in the Bravyi-Kitaev basis tells us whether or not the state of the quantum computer needs to acquire a phase change of $-1$ under the action of a creation or annihilation operator acting on orbital $j$. The parity of this set of qubits has the same parity as the set of orbitals with index less than $j$, and so we will  call this set of qubit indices the ``parity set" of index $j$, or $P(j)$. To determine the elements of $P(j)$, we consider the transformation from the Bravyi-Kitaev basis to the parity basis. From equation~(\ref{threeone}) we know that $p_i = \sum_j [ \pi_n]_{ij}\ f_j$. Given the inverse transformation matrix $\beta_n^{-1}$, it is also true that:
\begin{equation}
f_j = \sum_k [ \beta_n^{-1}]_{jk}\ b_k,
\end{equation}
\noindent and hence:
\begin{eqnarray}
p_i &= \sum_j [ \pi_n]_{ij}\ (\sum_k  [ \beta_n^{-1}]_{jk}\ b_k)\\
 &= \sum_k  [\pi_n \beta_n^{-1}]_{ik}\ b_k 
\end{eqnarray}

The matrix $\pi_n \beta_n^{-1}$ is the transformation matrix from the Bravyi-Kitaev basis to the parity basis. Therefore, the nonzero entries to the right of the main diagonal in row $i$ of the matrix $\pi_n \beta_n^{-1}$ give the indices of qubits in the Bravyi-Kitaev basis that can be used to compute the cumulative parity of orbitals with index less than $i$. An entry of 1 in row $i$, column $j$ of $\pi_n \beta_n^{-1}$ (where $j < i$, i.e. to the right of the main diagonal by our numbering) indicates that $j \in P(i)$:
\begin{equation}
\pi_8 \beta_8^{-1} =
\bordermatrix{~ & _7 & _6 & _5 & _4 & _3 & _2 & _1 & _0 \cr
                  _7 & 1 & 1 & 1 & 0 & 1 & 0 & 0 & 0 \cr
                  _6 & 0 & 1 & 1 & 0 & 1 & 0 & 0 & 0 \cr
                  _5 & 0 & 0 & 1& 1 & 1 & 0 & 0 & 0 \cr
                  _4 & 0 & 0 & 0 & 1 & 1 & 0 & 0 & 0 \cr
                  _3 & 0 & 0 & 0 & 0 & 1 & 1 & 1 & 0 \cr
                  _2 & 0 & 0 & 0 & 0 & 0 & 1 & 1 & 0 \cr
                  _1 & 0 & 0 & 0 & 0 & 0 & 0 & 1 & 1 \cr
                  _0 & 0 & 0 & 0 & 0 & 0 & 0 & 0 & 1 \cr}
                  \qquad
                  {\rm which~implies:}
                  \qquad
                  \left\{ 
  \begin{array}{l}
    P(7) = \{6,5,3\}\\
    P(6) = \{5,3\}\\
    P(5) = \{4,3\}\\
    P(4) = \{3\}\\
    P(3) = \{2,1\}\\
    P(2) = \{1\}\\
    P(1) = \{0\}\\
    P(0) = \emptyset \\
  \end{array} \right.
\end{equation}

\subsection{The update set}

For arbitrary $j$, we define the set of qubits (other than qubit $j$) that must be updated when the occupation of orbital $j$ changes. We call this set the ``update set" of index $j$, or $U(j)$. This is the set of qubits in the Bravyi-Kitaev basis that store a partial sum including orbital $j$. Any Bravyi-Kitaev qubit that stores a partial sum that includes occupation number $j$ is in $U(j)$. Since even indexed qubits store only the occupation of the corresponding orbital, update sets contain only odd indices. It is straightforward to determine the elements of $U(j)$ from the transformation matrix $\beta_n$ that maps bit strings in the occupation number basis to the Bravyi-Kitaev basis. The columns of this transformation matrix show which qubits in the Bravyi-Kitaev basis store a particular orbital, and so the nonzero entries in column $j$ above the main diagonal determine the qubits other than qubit $j$ that must be updated when the occupancy of orbital $j$ changes. These are the elements of the update set.

\begin{equation}
\beta_8 =
\bordermatrix{~ & f_7 & f_6 & f_5 & f_4 & f_3 & f_2 & f_1 & f_0   \cr
                  b_7 & 1 & 1 & 1 & 1 & 1 & 1 & 1 & 1 \cr
                  b_6 & 0 & 1 & 0 & 0 & 0 & 0 & 0 & 0 \cr
                  b_5 & 0 & 0 & 1 & 1 & 0 & 0 & 0 & 0 \cr
                  b_4 & 0 & 0 & 0 & 1 & 0 & 0 & 0 & 0 \cr
                  b_3 & 0 & 0 & 0 & 0 & 1 & 1 & 1 & 1 \cr
                  b_2 & 0 & 0 & 0 & 0 & 0 & 1 & 0 & 0 \cr
                  b_1 & 0 & 0 & 0 & 0 & 0 & 0 & 1 & 1 \cr
                  b_0 & 0 & 0 & 0 & 0 & 0 & 0 & 0 & 1 \cr}
                  \qquad
                  {\rm which~implies:}
                  \qquad
                  \left\{ 
  \begin{array}{l}
    U(7) = \emptyset\\
    U(6) = \{7\}\\
    U(5) = \{7\}\\
    U(4) = \{5,7\}\\
    U(3) = \{7\}\\
    U(2) = \{3,7\}\\
    U(1) = \{3,7\}\\
    U(0) = \{1,3,7\} \\
  \end{array} \right.
\end{equation}

It should be clear that update sets depend on the size of the basis used. For example, if 16 basis functions were used instead of the 8 used in the example above, all the update sets other than $U(7)$ would also include index 15.

\subsection{The flip set}
For arbitrary $j$, we need to know what set of Bravyi-Kitaev qubits determines whether qubit $j$ has the same parity or inverted parity with respect to orbital $j$.
We will call this set of Bravyi-Kitaev qubits the ``flip set" of $j$, or $F(j)$, because this set is responsible for whether $b_j$ has flipped parity with respect to $f_j$. This is the set that stores the parity of occupation numbers other than $f_j$ in the sum $b_j$. Since even-indexed qubits store only the orbital with the same index, the flip set of even indices is always the empty set. One can determine the elements of $F(j)$ by looking at the inverse transformation matrix $\beta_n^{-1}$ that maps bit strings in the Bravyi-Kitaev basis to the occupation number basis. The columns with nonzero entries to the right of the main diagonal in row $i$ of this inverse transformation matrix give the indices of the Bravyi-Kitaev qubits that together store the same set of orbitals as is stored by $\ket{b_i}$. These are the elements of the flip set.

\begin{equation}
\beta_8^{-1} =
\bordermatrix{~ & b_7 & b_6 & b_5 & b_4 & b_3 & b_2 & b_1 & b_0   \cr
                  f_7 & 1 & 1 & 1 & 0 & 1 & 0 & 0 & 0 \cr
                  f_6 & 0 & 1 & 0 & 0 & 0 & 0 & 0 & 0 \cr
                  f_5 & 0 & 0 & 1 & 1 & 0 & 0 & 0 & 0 \cr
                  f_4 & 0 & 0 & 0 & 1 & 0 & 0 & 0 & 0 \cr
                  f_3 & 0 & 0 & 0 & 0 & 1 & 1 & 1 & 0 \cr
                  f_2 & 0 & 0 & 0 & 0 & 0 & 1 & 0 & 0 \cr
                  f_1 & 0 & 0 & 0 & 0 & 0 & 0 & 1 & 1 \cr
                  f_0 & 0 & 0 & 0 & 0 & 0 & 0 & 0 & 1 \cr}
                  \qquad
                  {\rm which~implies:}
                  \qquad
                  \left\{ 
  \begin{array}{l}
    F(7) = \{6,5,3\} \\
    F(6) = \emptyset \\
    F(5) = \{4\}\\
    F(4) = \emptyset \\
    F(3) = \{2,1\}\\
    F(2) = \emptyset \\
    F(1) = \{0\} \\
    F(0) = \emptyset \\
  \end{array} \right.
\end{equation}
With these sets defined, we can derive the mapping from fermionic operators to qubit operators that is the equivalent of the Jordan-Wigner transformation in the Bravyi-Kitaev basis.

\section{The Bravyi-Kitaev transformation}\label{bravkit}
In this section we will give an explicit prescription, in terms of Pauli matrices, for representing the creation and
annihilation operators that act on the Bravyi-Kitaev basis states. Operating in this basis requires that we find the analogues to the qubit creation and annihilation operators 
($\hat{Q}^{\pm}$ in the occupation number basis, $\hat{\mathcal{P}}^{\pm}$ in the parity basis) as well as the
parity operator, $Z^{\rightarrow}_{i}$, and the update operator,  $X^{\leftarrow}_{i}$, in the Bravyi-Kitaev basis.
We will first define some notation.

For our purposes it is the parity of subsets of orbitals or qubits that matters, not the individual
occupation numbers or states of the qubits in the set. Thus, it is useful to define operators that project onto the subspace of the Hilbert space of the entire computer for which the subset of qubits with indices in $S$ has the parity selected for by the operator (even for ${\hat{E}}_S$, odd for ${\hat{O}}_S$). We can express these operators in terms of Pauli matrices as follows:
\begin{equation}\label{eq:5.2}
\hat{E}_S = \frac{1}{2}(\I + Z_S), \qquad {\hat{O}}_S = \frac{1}{2}(\I - Z_S),
\end{equation}
where $Z_S$ is shorthand for the $\sigma^z$ gate applied to all qubits in $S$. With this notation established, we will
next write equations for the qubit operators in the Bravyi-Kitaev basis that represent creation and annihilation operators acting on orbital $j$. To begin we will consider the case for which $j$ is even, because this will allow us to build intuition for the more difficult case for which $j$ is odd.

\subsection{Representing $a_j^{(\dag)}$ in the Bravyi-Kitaev basis for $j$ even}

In the case that $j$ is even, we should act with $\hat{Q}^\pm$ on qubit $j$, just as for the Jordan-Wigner transformation, because the Bravyi-Kitaev encoding stores orbitals with $j = 0 \pmod 2$ in the qubit with the same index. There are then two additional tasks that dictate how to represent the fermionic operators in the Bravyi-Kitaev basis: determining the parity of occupied orbitals with index less than $j$, and updating qubits with index greater than $j$ that store a partial sum that includes occupation number $j$.

The parity of the set of qubits in $P(j)$ is equal to that of the set of orbitals with index less than $j$. By analogy with the Jordan-Wigner transformation, we act with $\sigma^z$ on all qubits with indices in $P(j)$, that is, we apply the operator $Z_{P(j)}$. The number of qubits in $P(j)$ scales as $O(\log j) \leq O(\log n)$ \cite{BK}. 

Secondly, by analogy with the parity basis method, we also act with $\sigma^x$ on all qubits in the appropriate $U(j)$; that is, we apply the operator $X_{U(j)}$. This has the effect of updating all the qubits that store a set of orbitals including orbital $j$.  The size of $U(j)$ also scales like $O(\log n)$~\cite{BK}. To summarize: to represent $a^\dag_j$ or $a_j$ in the Bravyi-Kitaev basis, for $j$ even, we act with $\sigma^z$ on all qubits in $P(j)$, $\hat{Q}^{\pm}$ on qubit $j$, and with $\sigma^x$ on all qubits in $U(j)$:
\begin{eqnarray}
&a^\dag_j \equiv X_{U(j)} \otimes \hat{Q}^{+}_j \otimes Z_{P(j)} = \frac{1}{2}(X_{U(j)} \otimes X_j \otimes Z_{P(j)} - i X_{U(j)} \otimes Y_j \otimes Z_{P(j)}); \\
&a_j \equiv X_{U(j)} \otimes \hat{Q}^{-}_j \otimes Z_{P(j)}  = \frac{1}{2}(X_{U(j)} \otimes X_j \otimes Z_{P(j)} + i X_{U(j)} \otimes Y_j \otimes Z_{P(j)}).
\end{eqnarray}
In the next section, we will consider the case for which $j$ is odd.

\subsection{Representing $a_j^{(\dag)}$ in the Bravyi-Kitaev basis for $j$ odd}

To represent the creation or annihilation of a particle in orbital $j$ in the Bravyi-Kitaev basis, for $j$ even, we could simply act with
$\hat{Q}^{\pm}$ on qubit $j$ because that qubit stores only the occupation of orbital $j$. For $j$ odd, qubit $j$ stores a partial sum of
occupation numbers of orbitals including, but not limited to, orbital $j$. Thus, in this case the state of Bravyi-Kitaev qubit $j$ is either equal to the occupation of orbital $j$ (if the parity of the other orbitals that it stores is even), or opposite to that of orbital $j$ (if the parity of the other orbitals that it stores is 1). Thus, whether representing the creation or annihilation of a particle in orbital
$j$ requires that we act with
$\hat{Q}^+$ or $\hat{Q}^-$ on qubit $j$ in the Bravyi-Kitaev basis depends on the parity of all occupation numbers other than $f_j$ that are
included in the partial sum $b_j$ --- i.e. the parity of the flip set of index $j$. If the parity of the set of qubits with indices in $F(j)$ is even, then the creation or annihilation of a particle in orbital $j$ requires acting with $\hat{Q}^+$
or $\hat{Q}^-$, respectively, as usual. But if the parity of this set of qubits is odd, then the creation of a particle requires acting with $\hat{Q}^-$ and the annihilation of a particle requires acting with $\hat{Q}^+$. The Bravyi-Kitaev analogues to the qubit creation and annihilation operators are therefore:

\begin{equation}\label{eq:5.4}
\hat{\Pi}^{\pm}_j \equiv \hat{Q}^{\pm}_j \otimes \hat{E}_{F(j)} - \hat{Q}^{\mp}_j \otimes \hat{O}_{F(j)} = \frac{1}{2}(X_j \otimes Z_{F(j)} \mp i Y_j).
\end{equation}
\noindent The updating procedure in this case in which $j$ is odd works in exactly the same way as it does in the case that $j$ is even. In applying the parity operator, however, we need only consider the qubits that are in $P(j)$ but not in $F(j)$, because the relative sign in the $\hat{\Pi}^{\pm}_j$ operator implicitly calculates the parity of the subset of the parity set that is also in the flip set of index $j$. It is convenient to therefore introduce the new ``remainder set":
\begin{equation}
R(j) \equiv P(j) \setminus F(j).
\end{equation}
Thus, the fermionic creation and annihilation operators acting on orbital $j$ for $j$ odd are represented in the Bravyi-Kitaev basis as follows:
\begin{eqnarray}
&a^\dag_j \equiv X_{U(j)} \otimes \hat{\Pi}^{+}_j \otimes Z_{R(j)} = \frac{1}{2}(X_{U(j)} \otimes X_j \otimes Z_{P(j)} - i X_{U(j)} \otimes Y_j \otimes Z_{R(j)}); \\
&a_j \equiv X_{U(j)} \otimes \hat{\Pi}^{-}_j \otimes Z_{R(j)} =  \frac{1}{2}(X_{U(j)} \otimes X_j \otimes Z_{P(j)} + i X_{U(j)} \otimes Y_j \otimes Z_{R(j)}).
\end{eqnarray}

\noindent It is evident by inspection that the only difference in the algebraic form of the operators between the even- and odd-indexed cases is that the second term involves $Z_{P(j)}$ for the even case, but $Z_{R(j)}$ for the odd case. Therefore we define:

\begin{equation}
  \rho(j) \equiv \left\{ 
  \begin{array}{l l}
    P(j) & \quad {\rm if~}j{\rm~is~even;}\\
    R(j) & \quad {\rm if~}j{\rm~is~odd.}\\
  \end{array} \right.
\end{equation}

\noindent Now the fermionic creation and annihilation operators acting on arbitrary $j$ are represented in the Bravyi-Kitaev basis as:

\begin{eqnarray}
&a^\dag_j \equiv X_{U(j)} \otimes \hat{\Pi}^{+}_j \otimes Z_{R(j)} = \frac{1}{2}(X_{U(j)} \otimes X_j \otimes Z_{P(j)} - i X_{U(j)} \otimes Y_j \otimes Z_{\rho(j)}); \\
&a_j \equiv X_{U(j)} \otimes \hat{\Pi}^{-}_j \otimes Z_{R(j)} =  \frac{1}{2}(X_{U(j)} \otimes X_j \otimes Z_{P(j)} + i X_{U(j)} \otimes Y_j \otimes Z_{\rho(j)}).
\end{eqnarray}

\noindent These are useful basic results, but the operators that appear in the molecular electronic Hamiltonian are actually products of these creation and annihilation operators. In the next section, we derive general expressions for products of these second-quantized operators.

\section{Pauli representations of second-quantized operators in the Bravyi-Kitaev basis}\label{pauli}

In this Section we derive simplified algebraic expressions for classes of Hermitian second-quantized fermionic operators in the Bravyi-Kitaev basis. The five relevant classes of operators are summarized in Table~\ref{tab1}. We will give complete compact algebraic expressions for only the number operators and the Coulomb and exchange operators. It is not possible to give the algebraic form for the remaining three classes of operators without considering an impractical number of sub-cases, so we opt to give general expressions for products of the form $a_i^\dag a_j$, and show how to use these results to generate algebraic expressions for the remaining classes of operators.

\begin{table}[!h]
\begin{center}
\begin{tabular}{ |c | c| }
 \hline
 {\bf Operator} & {\bf Second quantized form} \\ 
 \hline
Number operator & $h_{ii}\  a^\dag_i a_i$ \\
\hline
Coulomb/exchange operators & $h_{ijji} \ a^\dag_i a^\dag_j a_j a_i$ \\
\hline
Excitation operator & $h_{ij} \ (a^\dag_i a_j + a^\dag_j a_i)$ \\ 
\hline
Number-excitation operator & $h_{ijjk} \ (a^\dag_i a^\dag_j a_j a_k + a^\dag_k a^\dag_j a_j a_i)$ \\ 
\hline
Double excitation operator & $h_{ijkl} \ (a^\dag_i a^\dag_j a_k a_l + a^\dag_l a^\dag_k a_j a_i)$ \\
\hline
\end{tabular}
\caption{The five classes of Hermitian second quantized operators that appear in electronic Hamiltonians. In general the overlap integrals $h_{ij}$ and $h_{ijkl}$ may be complex.\label{tab1}}
\end{center}
\end{table}

\subsection{Number operators: $h_{ii}\ a^\dag_i a_i$ }\label{sixone}
The number operators are of the form $h_{ii}\ a^\dag_i a_i$ and have eigenvalues corresponding to the occupation number of orbital $i$. We would like to find a simplified expression for this class of operators in the Bravyi-Kitaev basis. 

Given the results of Section~\ref{bravkit}, we can write the following:
\begin{eqnarray}
a^\dag_i a_i = \ &\frac{1}{2}(X_{U(i)} \otimes X_i \otimes Z_{P(i)} - i X_{U(i)} \otimes Y_i \otimes Z_{\rho(i)}) \\
\times &\frac{1}{2}(X_{U(i)} \otimes X_i \otimes Z_{P(i)} + i X_{U(i)} \otimes Y_i \otimes Z_{\rho(i)}). \nonumber
\end{eqnarray}
Given that $\sigma^x \sigma^x = \sigma^y \sigma^y = \sigma^z \sigma^z = \I$, it follows that $(X_S)^2 = (Y_S)^2 = (Z_S)^2 = \I$. We are left with:
\begin{eqnarray}
a^\dag_i a_i &= \frac{1}{4}[\I + i (X_i Y_i) \otimes Z_{P(i)\setminus \rho(i)} - i (Y_i X_i) \otimes Z_{P(i)\setminus \rho(i)} + \I]  \\
 &= \frac{1}{2}(\I - Z_i \otimes Z_{P(i)\setminus \rho(i)}).
\end{eqnarray}
Now, when $i$ is even, $\rho(i) = P(i)$, and so $P(i) \setminus \rho(i)  = \emptyset$. When $i$ is odd, $\rho(i) = R(i)$, and so $P(i) \setminus \rho(i)  = F(i)$. Conveniently, $F(i) = \emptyset$ for $i$ even, so if we define the following:
\begin{equation}
\underline{F(i)} \equiv F(i) \cup \{i\},
\end{equation}
then we can represent the number operators for arbitrary $i$  (even or odd) as follows:
\begin{equation}
a^\dag_i a_i = \frac{1}{2}(\I - Z_{\underline{F(i)}}).
\end{equation}
In the next section we consider the Coulomb and exchange operators.

\subsection{Coulomb and exchange operators: $h_{ijji}\ a^\dag_i a_j^\dag a_j a_i$}

The Coulomb operators are of the form $a^\dag_i a_j^\dag a_j a_i$, while the exchange operators are of the form $a^\dag_i a_j^\dag a_i a_j = - a^\dag_i a_j^\dag a_j a_i$. Since these two kinds of operators can be grouped together algebraically, we consider them as one case. The fermionic anti commutation relations ensure that $a^\dag_i a_j^\dag a_j a_i = - a^\dag_i a_j^\dag a_i a_j  =  (a^\dag_i a_i)(a_j^\dag a_j)$. Thus, we can consider the Coulomb and exchange operators as a product of two number operators. With the result from Section~\ref{sixone}, we can write the following:

\begin{eqnarray}
 a^\dag_i a_j^\dag a_j a_i = \ &\frac{1}{2}(\I - Z_{\underline{F(i)}}) \times \frac{1}{2}(\I - Z_{\underline{F(j)}})  \\
 = \ &\frac{1}{4}(\I - Z_{\underline{F(i)}} - Z_{\underline{F(j)}} + Z_{\underline{F(i)}} Z_{\underline{F(j)}}).
 \end{eqnarray}
Any overlap between supp($Z_{\underline{F(i)}}$) and supp($Z_{\underline{F(j)}}$), where supp($\hat{O}$) is the support of the operator $\hat{O}$, i.e. those tensor factors on which it acts nontrivially, will result in the local product $\sigma^z \sigma^z = \I$. Thus, we only actually need to act with $\sigma^z$ on the union of $\underline{F(i)}$ and $\underline{F(j)}$ minus their intersection, i.e. the symmetric difference of these two sets. Thus we define the following notation:
\begin{equation}
\underline{F_{ij}} \equiv \underline{F(i)} \bigtriangleup \underline{F(j)} = (\underline{F(i)} \cup \underline{F(j)}) \setminus (\underline{F(i)} \cap \underline{F(j)}).
\end{equation}
We can then give the algebraic expression for the Coulomb and exchange operators:
\begin{equation}
a^\dag_i a_j^\dag a_j a_i = \frac{1}{4}(\I - Z_{\underline{F(i)}} - Z_{\underline{F(j)}} + Z_{\underline{F_{ij}}}).
\end{equation}
In the next section we consider general products of the form $a_i^\dag a_j$.

\subsection{Products of the form $a^\dag_i a_j$}

We can assume without loss of generality that $i < j$. The algebraic form for products of this kind depends on the parity of the indices. There are four cases and we will work through the first case in detail, and simply present the results for the other cases.

Using the result of Section~\ref{bravkit}, we obtain the following when $i$ and $j$ are even:
\begin{eqnarray}\label{sixeight}
a_i^\dag a_j = \ &\frac{1}{2}(X_{U(i)} \otimes X_i \otimes Z_{P(i)} - i X_{U(i)} \otimes Y_i \otimes Z_{P(i)}) \\
\times \ &\frac{1}{2}(X_{U(j)} \otimes X_j \otimes Z_{P(j)} + i X_{U(j)} \otimes Y_j \otimes Z_{P(j)}). \nonumber
\end{eqnarray}

For each of the four terms resulting from multiplying out the operators in equation~(\ref{sixeight}) above, we must consider what products of local qubit operators can result. There are three potential sources of local qubit operator products: overlap between the update set of qubit $i$ and the update set of qubit $j$, overlap between the update set of qubit $i$ and the parity set of qubit $j$, and overlap between the parity set of qubit $i$ and the parity set of qubit $j$. Any overlap between the update sets of qubits $i$ and $j$ will result in the local product $\sigma^x  \sigma^x = \I$; any overlap between the update set of qubit $i$ and parity set of qubit $j$ will result in the local product $\pm i 
\sigma^y$; and any overlap in the parity sets of qubits $i$ and $j$ will result in the local product $\sigma^z \sigma^z = \I$. Thus we define the following sets:
\begin{equation}
U_{ij} \equiv U(i) \bigtriangleup U(j), \quad \quad \alpha_{ij} \equiv U(i) \cap P(j), \quad \quad P_{ij}^0 \equiv P(i) \bigtriangleup P(j).
\end{equation}
Note that in the case that $i$ and $j$ are even, we do not need to consider the possibility that $j \in U(i)$ because $U(i)$ contains only odd elements. Similarly, we do not need to consider the possibility that $i \in P(j)$, because $P(j)$ for $j$ even contains only odd elements.

As an example, we will show how to use the sets defined above to simplify the term $(X_{U(i)} \otimes X_i \otimes Z_{P(i)})(X_{U(j)} \otimes X_j \otimes Z_{P(j)})$. For this term, we need only apply $\sigma^x$ to the set of qubits $U_{ij} \setminus \alpha_{ij} \cup \{i,j\}$, $\sigma^y$ to the qubit with index in $\alpha_{ij}$ (which set in general has at most 1 element, and in the case that $i$ and $j$ are even always contains 1 element), and $\sigma^z$ to the qubits in the set $P_{ij}^0 \setminus \alpha_{ij}$. Thus, this term simplifies to:
\begin{equation}
(X_{U(i)} \otimes X_i \otimes Z_{P(i)})(X_{U(j)} \otimes X_j \otimes Z_{P(j)}) = -i \ X_{U_{ij} \setminus \alpha_{ij} \cup \{i,j\}}  Y_{\alpha_{ij}}  Z_{P_{ij}^0 \setminus \alpha_{ij}}.
\end{equation}
Using the same reasoning for the other terms, we arrive at the following result:
 \begin{equation}
a_i^\dag a_j = \frac{1}{4} X_{U_{ij} \setminus \alpha_{ij}}  Y_{\alpha_{ij}} Z_{P_{ij}^0 \setminus \alpha_{ij}} [Y_j X_i - X_j Y_i -i (X_j X_i + Y_j Y_i)].
\end{equation}
This is our result for the case that $i$ and $j$ are even. The algebraic expressions for the other cases can be derived in the same manner, with the added complication that the expression for the product $a_i^\dag a_j$ varies, depending on if $i \in P(j)$ and/or $j \in U(i)$. This complication results in a proliferation of sub-cases: two for the case that $i$ is odd and $j$ is even, three for the case that $i$ is even and $j$ is odd, and four for the case that $i$ and $j$ are odd. The only additional sets we need to define are the analogs of $P_{ij}^0$ for when one or both of the indices are odd:
\begin{equation}
P_{ij}^1 \equiv P(i) \bigtriangleup R(j), \quad \quad P_{ij}^2 \equiv R(i) \bigtriangleup P(j), \quad \quad P_{ij}^3 \equiv R(i) \bigtriangleup R(j).
\end{equation}
The results for all cases are summarized below in Table~\ref{tab2}. In the following sub-sections we show how to use the contents of Table~\ref{tab2} to generate algebraic expressions for the excitation operators, the number-excitation operators, and the double-excitation operators.

\subsection{Excitation operators: $h_{ij} \ (a_i^\dag a_j + a_j^\dag a_i)$}
Providing for the possibility that the integral $h_{ij}$ is complex, we can write:
\begin{equation}
h_{ij} \ (a_i^\dag a_j + a_j^\dag a_i) = \Re\{h_{ij}\} (a_i^\dag a_j + a_j^\dag a_i) + \Im\{h_{ij}\} (a_i^\dag a_j - a_j^\dag a_i).
\end{equation}
Applying this to the case when $i$ and $j$ are even, we find the following:
\begin{eqnarray}
h_{ij} \ (a_i^\dag a_j + a_j^\dag a_i) = \frac{1}{2} X_{U_{ij} \setminus \alpha_{ij}}\  Y_{\alpha_{ij}}\ Z_{P_{ij}^0 \setminus \alpha_{ij}} [&\Re\{h_{ij}\}(Y_j X_i - X_j Y_i) \\
 + &\Im\{h_{ij}\} (X_j X_i + Y_j Y_i)]. \nonumber
\end{eqnarray}
Similar expressions for other cases are easily generated by taking the appropriate form of $a_i^\dag a_j$ from Table~\ref{tab2}.

\subsection{Number-excitation operators: $h_{ijjk} \ (a^\dag_i a^\dag_j a_j a_k + a^\dag_k a^\dag_j a_j a_i)$}

Due to the fermionic anti-commutation relations, the following is true:
\begin{equation}
a^\dag_i a^\dag_j a_j a_k + a^\dag_k a^\dag_j a_j a_i = (a_i^\dag a_k + a_k^\dag a_i) (a_j^\dag a_j).
\end{equation}
We see that this is simply a product of an excitation operator and a number operator. We have previously given algebraic expressions for both of these classes of operators, so it is not difficult to combine them for an expression for the number-excitation operators. Let us consider the example when $i$ and $k$ are even. Then we have the following:
\begin{eqnarray}
h_{ijjk} \ (a_i^\dag a_k + a_k^\dag a_i) a_j^\dag a_j = &\frac{1}{2} X_{U_{ik} \setminus \alpha_{ik}}\  Y_{\alpha_{ik}}\ Z_{P_{ik}^0 \setminus \alpha_{ik}} [\Re\{h_{ijjk}\}(Y_k X_i - X_k Y_i)  \\
&+ \Im\{h_{ijjk}\} (X_k X_i + Y_k Y_i)] \times \frac{1}{2}(\I - Z_{\underline{F(j)}}). \nonumber
\end{eqnarray}
To simplify, all we need to consider is the intersection between $\underline{F(j)}$ and the support of $(a_i^\dag a_k + a_k^\dag a_i)$. In this case the support of the excitation operator is $U_{ik} \cup \alpha_{ik} \cup P_{ik}^0 \cup \{i,k\}$. The form of the simplification will vary depending on these sets, but the process of reducing local operator products by exploiting the relationship between the three Pauli matrices is unchanged. In the cases when $i$ and $k$ are not both even, all that changes is the form of the excitation operator from Table~\ref{tab2} that must be used.

\subsection{Double-excitation operators: $h_{ijkl} \ (a^\dag_i a^\dag_j a_k a_l + a^\dag_l a^\dag_k a_j a_i)$}
The double-excitation operators involve four distinct indices, and are obviously the most algebraically complicated class of operators we are considering. The impractical number of sub-cases depending on the specific combination of indices $i,j,k,l$ means that we only outline the procedure for deriving algebraic expressions for this class of operators.
The fermionic commutation relations ensure that the following is true:
\begin{equation}
(a^\dag_i a^\dag_j a_k a_l + a^\dag_l a^\dag_k a_j a_i) = (a^\dag_i a_l) (a^\dag_j a_k) + (a^\dag_l a_i) (a^\dag_k a_j).
\end{equation}
Allowing for the integral $h_{ijkl}$ to be complex, we can write:
\begin{eqnarray}
h_{ijkl} \ (a^\dag_i a^\dag_j a_k a_l + a^\dag_l a^\dag_k a_j a_i) = [&\Re\{h_{ijkl}\}(a^\dag_i a_l a^\dag_j a_k + a^\dag_l a_i a^\dag_k a_j)  \\ + &\Im\{h_{ijkl}\}\ (a^\dag_i a_l a^\dag_j a_k - a^\dag_l a_i a^\dag_k a_j)]. \nonumber
\end{eqnarray}
\noindent Since $(a^\dag_i a_l a_j^\dag a_k)^\dag = a^\dag_l a_i a^\dag_k a_j$, we can simply consider the algebraic expression for the product of two operators of the form $a_i^\dag a_j$ as given in Table~\ref{tab2}, and then add or subtract it to its Hermitian conjugate. Each of the operators $a^\dag_i a_l$ and $a_j^\dag a_k$ will fit into one of the ten cases presented in Table~\ref{tab2}. In multiplying out the algebraic expressions for these two products, what is important is the set \{supp($a^\dag_i a_l$) $\cap$ supp($a_j^\dag a_k$)\}. Any qubits in this set will have a product of local operators acting on it which must be simplified. 

\begin{table}
\begin{tabular}{|c|c|c|c|c|}
\hline
Index parity & & Conditions & & Algebraic expression for $a_i^\dag a_j$ \\
\hline
 & $i \in P(j)$ & $j \in U(i)$ & $|\alpha_{ij}| $& \\
\hline
\multirow{1}{*}{$i,j$ even} & No & No & 1 & $\frac{1}{4} X_{U_{ij} \setminus \alpha_{ij}}\  Y_{\alpha_{ij}}\ Z_{P_{ij}^0 \setminus \alpha_{ij}} [Y_j X_i - X_j Y_i -i (X_j X_i + Y_j Y_i)]$ \\ 
\hline
\multirow{3}{*}{$i$ odd, $j$ even}& No & No & 1  & $\frac{1}{4} X_{U_{ij} \setminus \alpha_{ij}}\  Y_{\alpha_{ij}}\ \overline{Z}_{\alpha_{ij}}\ [(Y_j X_i - i X_j X_i)\ Z_{P_{ij}^0}  - (X_j Y_i + iY_j Y_i)\ Z_{P_{ij}^2} ] $ \\
& & & & \\
   & Yes & No & 0 &  $\frac{1}{4} X_{U_{ij}}\ \overline{Z}_i \ [(Y_j Y_i - i X_j \overline{X}_i Y_i)\ Z_{P_{ij}^0}  + (X_j X_i + iY_j X_i)\ Z_{P_{ij}^2} ]$ \\ 
\hline
\multirow{5}{*}{$i$ even, $j$ odd} & No & No & 1  & $\frac{1}{4} X_{U_{ij} \setminus \alpha_{ij}}\  Y_{\alpha_{ij}}\ \overline{Z}_{\alpha_{ij}}\ [- (X_j Y_i + i X_j X_i)\ Z_{P_{ij}^0}  + (Y_j X_i - iY_j Y_i)\ Z_{P_{ij}^1} ]$ \\
& & & & \\
      & No & Yes & 1  & $\frac{1}{4} X_{U_{ij} \setminus j}\  [-\overline{X}_{\alpha_{ij}}(Y_i - i X_i)\ Y_{\alpha_{ij}}\ Z_{P_{ij}^0 \setminus \alpha_{ij}}  + (i Y_i - X_i)\ Z_{P_{ij}^1 \cup j} ]$ \\
      & & & & \\
      & Yes & Yes & 0  & $\frac{1}{4} X_{U_{ij} \setminus j}\  [ (X_i - i Y_i)  + (i Y_i - X_i)\ Z_{P_{ij}^1 \cup j} ]$ \\ \midrule
\hline
\multirow{7}{*}{$i,j$ odd} & No & No & 1 & $\frac{1}{4} X_{U_{ij} \setminus \alpha_{ij}}\ Y_{\alpha_{ij}} \overline{Z}_{\alpha_{ij}}\ [ -i X_j X_i Z_{P_{ij}^0} + Y_j X_i Z_{P_{ij}^1} - X_j Y_i Z_{P_{ij}^2} - i Y_j Y_i Z_{P_{ij}^3}]$ \\
& & & & \\
  & Yes & No & 0 & $\frac{1}{4} X_{U_{ij}}\ \overline{Z}_i [ (-i X_j Y_i Z_{P_{ij}^0} + Y_j Y_i Z_{P_{ij}^1}) + X_j X_i Z_{P_{ij}^2} + i Y_j X_i Z_{P_{ij}^3}]$ \\
  & & & & \\
  & No & Yes & 1 & $\frac{1}{4} X_{U_{ij} \setminus j}\ [ -\overline{X}_{\alpha_{ij}}(Y_i Z_{P_{ij}^2} + i X_i Z_{P_{ij}^0}) Y_{\alpha_{ij}} \overline{Z}_{\alpha_{ij}} - (X_i Z_{P_{ij}^1} - i Y_i Z_{P_{ij}^3}) Z_j ]$ \\
  & & & & \\
  & Yes & Yes &0 & $\frac{1}{4} X_{U_{ij} \setminus j}\ [ \overline{Z}_i (-i Y_i Z_{P_{ij}^0} + X_i Z_{P_{ij}^2}) + Z_j (-X_i Z_{P_{ij}^1} + i Y_i Z_{P_{ij}^3})]$ \\
\hline
\end{tabular}
\caption{\label{tab2}The algebraic expressions for general products of the form $a_i^\dag a_j$ in the Bravyi-Kitaev basis. These expressions vary in form depending on the parity of the indices $i$ and $j$, as well as on the overlaps between the parity and update sets of the indices. The notation $\overline{O}_S$ is shorthand to indicate that the operator $O$ does {\em not} operate on the qubits in the set $S$ (i.e. $Z_{P_{ij}^0} \overline{Z}_j= Z_{P_{ij}^0 \setminus j}$).}
\end{table}

\section{The molecular electronic Hamiltonian for the hydrogen molecule in the Bravyi-Kitaev basis}\label{molham}

The molecular electronic Hamiltonian~(\ref{molhameq}) may be divided into one and two-electron terms:
\begin{equation}
\hat{H}=\sum_{i,j}h_{ij}a^\dag_i a_j +\frac{1}{2}\sum_{i,j,k,l} h_{ijkl} a^\dag_i a^\dag_j a_k a_l = \hat{H}^{(1)} + \hat{H}^{(2)}.
\end{equation}

We treat molecular hydrogen in a minimal basis, so the sums above run over the four spin orbitals defined above. These spin orbitals will be indexed 0 through 3, as will be the fermionic creation and annihilation operators. We derive the simplified expressions for the individual terms of this Hamiltonian in the Bravyi-Kitaev basis. The overlap integrals $h_{ij}$ and $h_{ijkl}$ for $0 \leq i \leq 3$ are given in Table~\ref{tab3}. These are the same as were used in \cite{James} and were calculated using a restricted Hartree-Fock calculation in the PyQuante quantum chemistry package \cite{Muller}. With these integrals and the algebraic expressions for second quantized operators given in Section~\ref{pauli}, we can express the molecular electronic Hamiltonian for H$_2$ as a sum of products of Pauli matrices. In the next two subsections we consider the one- and two-electron Hamiltonians separately.

\begin{table}[h!]
\begin{center}
\begin{tabular}{ |c| c| }
\hline
Integrals & Value (a.u.) \\
\hline
$h_{00} = h_{11}$ & $-1.252477$ \\
\hline
$h_{22} = h_{33}$& $-0.475934$\\
\hline
$h_{0110} = h_{1001}$ & $\ 0.674493$\\
\hline
$h_{2332} = h_{3223}$ & $\ 0.697397$\\
\hline
$\quad h_{0220} = h_{0330} = h_{1221} = h_{1331}$ &\multirow{2}{*}{$0.663472$}\\
$= h_{2002} = h_{3003} = h_{2112} = h_{3113}$ & \\
\hline
$h_{0202} = h_{1313} = h_{2130} = h_{2310} = h_{0312} = h_{0132}$ & $0.181287$\\
\hline
\end{tabular}
\caption{\label{tab3}The overlap integrals for molecular hydrogen in a minimal basis. The integrals were obtained through a restricted Hartree-Fock calculation in the PyQuante quantum chemistry package at an internuclear separation of $1.401000$ atomic units ($7.414 \times 10^{-11}$ m).}
\end{center}
\end{table}

\subsection{The Bravyi-Kitaev Pauli representation of $\hat{H}^{(1)}$}\label{sevenone}

We can write the one-electron terms in the Hamiltonian as:
\begin{equation}
\hat{H}^{(1)} = h_{00}a_0^\dag a_0 + h_{11}a_1^\dag a_1 + h_{22}a_2^\dag a_2 + h_{33}a_3^\dag a_3.
\end{equation}
Using the expressions for number operators derived in Section~\ref{bravkit}, we know that in the Bravyi-Kitaev basis, these operators are:
\begin{eqnarray}
&a_0^\dag a_0 =  \frac{1}{2}(\I - \sigma_0^z); \\
&a_1^\dag a_1 = \frac{1}{2}(\I - \sigma_1^z \sigma_0^z); \\
&a_2^\dag a_2 = \frac{1}{2}(\I - \sigma_2^z); \\
&a_3^\dag a_3 =  \frac{1}{2}(\I - \sigma_3^z \sigma_2^z \sigma_1^z).
\end{eqnarray}
We now proceed to the simulation of $\hat{H}^{(2)}$.

\subsection{The Bravyi-Kitaev Pauli representation of $\hat{H}^{(2)}$}

Following the work of Whitfield {\it et al}.\ \cite{James}, $\hat{H}^{(2)}$ simplifies to the following expression for molecular hydrogen in a minimal basis:
\begin{eqnarray}
\hat{H}^{(2)} = h_{0110}a_0^\dag a_1^\dag a_1 a_0 + h_{2332}a_2^\dag a_3^\dag a_3 a_2 &+ h_{0330}a_0^\dag a_3^\dag a_3 a_0 + h_{1221}a_1^\dag a_2^\dag a_2 a_1   \\ 
\ + (h_{0220} - h_{0202})a_0^\dag a_2^\dag a_2 a_0 + (h_{1331} - h_{1313}) &a_1^\dag a_3^\dag a_3 a_1 + h_{0132}(a_0^\dag a_1^\dag a_3 a_2 + a_2^\dag a_3^\dag a_1 a_0) \nonumber \\
+ h_{0312}(a_0^\dag a_3^\dag a_1 a_2 &+ a_2^\dag a_1^\dag a_3 a_0). \nonumber
\end{eqnarray}
This term in the Hamiltonian is made up of six Coulomb/exchange operators and two double-excitation operators. Using Section~\ref{pauli}, it is easy to give algebraic expressions for the Coulomb and exchange operators:
\begin{eqnarray}
&a_0^\dag a_1^\dag a_1 a_0 = \frac{1}{4}(\I - \sigma_0^z - \sigma_1^z \sigma_0^z + \sigma_1^z); \\
&a_2^\dag a_3^\dag a_3 a_2 = \frac{1}{4}(\I - \sigma_2^z - \sigma_3^z \sigma_2^z \sigma_1^z + \sigma_3^z \sigma_1^z); \\
&a_0^\dag a_3^\dag a_3 a_0 = \frac{1}{4}(\I - \sigma_0^z - \sigma_3^z \sigma_2^z \sigma_1^z + \sigma_3^z \sigma_2^z \sigma_1^z \sigma_0^z ); \\
&a_1^\dag a_2^\dag a_2 a_1 = \frac{1}{4}(\I - \sigma_2^z - \sigma_1^z \sigma_0^z + \sigma_2^z \sigma_1^z \sigma_0^z); \\
&a_0^\dag a_2^\dag a_2 a_0 = \frac{1}{4}(\I - \sigma_2^z - \sigma_0^z + \sigma_2^z \sigma_0^z); \\
&a_1^\dag a_3^\dag a_3 a_1 = \frac{1}{4}(\I - \sigma_3^z \sigma_2^z \sigma_1^z - \sigma_1^z \sigma_0^z + \sigma_3^z \sigma_2^z \sigma_0^z).
\end{eqnarray}
The two double-excitation operators are somewhat more complicated. As an example, we will derive the Pauli representation of $h_{0312}(a_0^\dag a_3^\dag a_1 a_2 + a_2^\dag a_1^\dag a_3 a_0)$. Following in Section~\ref{pauli}, we  consider $a_0^\dag a_3^\dag a_1 a_2$ as $(a_0^\dag a_2) (a_3^\dag a_1)$, a product of two operators of the form $a_i^\dag a_j$. The term $a_0^\dag a_2$ is of the type when $i$ and $j$ are both even, while the term $a_1^\dag a_3$ is of the type when $i$ and $j$ are odd, and $i \in P(j)$, $j \in U(i)$, and $|\alpha_{ij}| = 0$. Using the appropriate expressions from Table~\ref{tab2}, we find the following:
\begin{eqnarray}
& a_0^\dag a_2 = \frac{1}{4} (\sigma^y_2 \sigma^y_1 \sigma^x_0 - \sigma^x_2 \sigma^y_1 \sigma^y_0 - i \sigma^x_2 \sigma^y_1 \sigma^x_0 -i \sigma^y_2 \sigma^y_1 \sigma^y_0); \\
& a_1^\dag a_3 = \frac{1}{4} (-i \sigma^z_2 \sigma^y_1 \sigma^z_0 + \sigma^z_2 \sigma^x_1 - \sigma^z_3 \sigma^x_1 \sigma^z_0 + i \sigma^z_3 \sigma^y_1).
\end{eqnarray}

\noindent Now we note that supp($a_0^\dag a_2$) $\cap$ supp($a_1^\dag a_3 $) $= \{2,1,0\}$, and so we must expect to simplify local operator products on qubits with these indices. Taking the product, we find the following:
\begin{eqnarray}
a_0^\dag a_2 a_1^\dag a_3 = \frac{1}{16} (&\sigma^x_2 \sigma^x_0 - i \sigma^x_2 \sigma^y_0 + \sigma^x_2 \sigma^z_1 \sigma^x_0 - i \sigma^x_2 \sigma^z_1 \sigma^y_0 \\
 + &i \sigma^y_2 \sigma^x_0 + \sigma^y_2 \sigma^y_0 + i \sigma^y_2 \sigma^z_1 \sigma^x_0 + \sigma^y_2 \sigma^z_1 \sigma^y_0  \nonumber \\
 + &\sigma^z_3 \sigma^x_2 \sigma^x_0 - i \sigma^z_3 \sigma^x_2 \sigma^y_0 + \sigma^z_3 \sigma^x_2 \sigma^z_1 \sigma^x_0 - i \sigma^z_3 \sigma^x_2 \sigma^z_1 \sigma^y_0 \nonumber \\
 + &i \sigma^z_3 \sigma^y_2 \sigma^x_0 + \sigma^z_3 \sigma^y_2 \sigma^y_0 + i \sigma^z_3 \sigma^y_2 \sigma^z_1 \sigma^x_0 + \sigma^z_3 \sigma^y_2 \sigma^z_1 \sigma^y_0). \nonumber
\end{eqnarray}

Since the integral $h_{0132}$ is real, we can simply add the above result to its Hermitian conjugate to find the expression for the double-excitation operator. Repeating the above procedure for the second double excitation operator, we arrive at the following results:
\begin{eqnarray}
a_0^\dag a_3^\dag a_1 a_2 + a_2^\dag a_1^\dag a_3 a_0 = \frac{1}{8}(&- \sigma_{2}^x\sigma_{0}^x\ + \sigma_{2}^x\sigma_{1}^z\sigma_{0}^x\ - \sigma_{2}^y\sigma_{0}^y\ + \sigma_{2}^y\sigma_{1}^z\sigma_{0}^y\ - \sigma_{3}^z\sigma_{2}^x\sigma_{0}^x\ \\
& + \sigma_{3}^z\sigma_{2}^x\sigma_{1}^z\sigma_{0}^x\ - \sigma_{3}^z\sigma_{2}^y\sigma_{0}^y\ + \sigma_{3}^z\sigma_{2}^y\sigma_{1}^z\sigma_{0}^y); \nonumber \\
a_0^\dag a_1^\dag a_3 a_2 + a_2^\dag a_3^\dag a_1 a_0 = \frac{1}{8}(& \sigma_{2}^x\sigma_{0}^x\ + \sigma_{2}^x\sigma_{1}^z\sigma_{0}^x\ + \sigma_{2}^y\sigma_{0}^y\ + \sigma_{2}^y\sigma_{1}^z\sigma_{0}^y\ + \sigma_{3}^z\sigma_{2}^x\sigma_{0}^x\  \\
& + \sigma_{3}^z\sigma_{2}^x\sigma_{1}^z\sigma_{0}^x\ + \sigma_{3}^z\sigma_{2}^y\sigma_{0}^y\ + \sigma_{3}^z\sigma_{2}^y\sigma_{1}^z\sigma_{0}^y). \nonumber
\end{eqnarray}
Thus, using the integrals from Table~\ref{tab3} and the Pauli expressions for the number operators derived in Section~\ref{sevenone}, as well as the Coulomb/exchange operators and the double-excitation operators derived in this section, we can represent the molecular electronic Hamiltonian for the hydrogen molecule as a sum of products of Pauli matrices in the Bravyi-Kitaev basis:
\begin{eqnarray}\label{svnnn}
\hat{H}_{BK} =\ &-0.81261\ \I+0.171201\ \sigma_{0}^z\ +0.16862325\ \sigma_{1}^z\ -0.2227965\ \sigma_{2}^z\ +0.171201\ \sigma_{1}^z \sigma_{0}^z\  \nonumber \\
&+0.12054625\ \sigma_{2}^z\sigma_{0}^z\ +0.17434925\ \sigma_{3}^z\sigma_{1}^z\ +0.04532175\ \sigma_{2}^x\sigma_{1}^z\sigma_{0}^x\ +0.04532175\ \sigma_{2}^y\sigma_{1}^z\sigma_{0}^y\    \nonumber \\
&+0.165868\ \sigma_{2}^z\sigma_{1}^z\sigma_{0}^z\ +0.12054625\ \sigma_{3}^z\sigma_{2}^z\sigma_{0}^z\ -0.2227965\ \sigma_{3}^z\sigma_{2}^z\sigma_{1}^z\  \nonumber \\
&+0.04532175\ \sigma_{3}^z\sigma_{2}^x\sigma_{1}^z\sigma_{0}^x\ +0.04532175\ \sigma_{3}^z\sigma_{2}^y\sigma_{1}^z\sigma_{0}^y\ +0.165868\ \sigma_{3}^z\sigma_{2}^z\sigma_{1}^z\sigma_{0}^z. 
\end{eqnarray}
This Hamiltonian is isospectral to the Jordan-Wigner derived Hamiltonian \cite{James}:
\begin{eqnarray}\label{svntwn}
\hat{H}_{JW} =\ &-0.81261\ \I+0.171201\ \sigma_{0}^z\ +0.171201\ \sigma_{1}^z\ -0.2227965\ \sigma_{2}^z\ -0.2227965\ \sigma_{3}^z\  \nonumber \\
&+0.16862325\ \sigma_{1}^z\sigma_{0}^z\ +0.12054625\ \sigma_{2}^z\sigma_{0}^z\ +0.165868\ \sigma_{2}^z\sigma_{1}^z\ +0.165868\ \sigma_{3}^z\sigma_{0}^z\ \nonumber \\
&+0.12054625\ \sigma_{3}^z\sigma_{1}^z\ +0.17434925\ \sigma_{3}^z\sigma_{2}^z\ -0.04532175\ \sigma_{3}^x\sigma_{2}^x\sigma_{1}^y\sigma_{0}^y\ \nonumber \\
&+0.04532175\ \sigma_{3}^x\sigma_{2}^y\sigma_{1}^y\sigma_{0}^x\ +0.04532175\ \sigma_{3}^y\sigma_{2}^x\sigma_{1}^x\sigma_{0}^y\ -0.04532175\ \sigma_{3}^y\sigma_{2}^y\sigma_{1}^x\sigma_{0}^x.
\end{eqnarray}
Writing the electronic Hamiltonians in the form of equations~(\ref{svnnn}) and~(\ref{svntwn}) allows for a comparison of the computational resources required to simulate them on a quantum computer. Not all tensor products of Pauli matrices that appear in these Hamiltonians commute with one another, so exponentiating them requires the use of a Trotter approximation. The next section details the Trotterization process for the Hamiltonian in the Bravyi-Kitaev basis.

\section{Trotterization}\label{trott}
Ideally, one could simulate the propagator $e^{-i \hat{H} t}$, where $\hat{H} = \sum_{k} h_k$, by sequentially exponentiating the individual terms $h_k$ on a quantum simulator. However, $e^{-i \hat{H} t} = \prod{e^{-i h_k t}}$ only in the case that the set of $h_k$ all mutually commute. Both the Bravyi-Kitaev and Jordan-Wigner Hamiltonians contain terms that do not commute with one another, and so a Suzuki-Trotter approximation must be used. The first four orders of Suzuki-Trotter formulae are \cite{QA}:
\begin{equation} e^{(A+B)t} \approx (e^{At/n}e^{Bt/n})^n+O(t \Delta t);
\end{equation}
\begin{equation} e^{(A+B)t} \approx (e^{At/2n}e^{Bt/n}e^{At/2n})^{n}+O(t (\Delta t)^2);
\end{equation}
\begin{equation} e^{(A+B)t} \approx (e^{\frac{7}{24}At/n}e^{\frac{2}{3}Bt/n}e^{\frac{3}{4}At/n}e^{\frac{-2}{3}Bt/n}e^{\frac{-1}{24}At/n}e^{Bt/n})^{n}+O(t (\Delta t)^3);
\end{equation}
\begin{equation} e^{(A+B)t} \approx (\prod{_{i=1}^5}e^{p_i At/2n}e^{p_i Bt/n}e^{p_i At/2n})^n+O(t (\Delta t)^4),
\end{equation}

\noindent where in the 4th order equation, the constants are given by:
\begin{equation}
p_1 =p_2 =p_4 =p_5 = \frac{1}{4-4^{1/3}}, \qquad p_3=1-4p_1.
\end{equation}

\begin{figure}[!h]
\centerline{ \includegraphics[scale = .35]{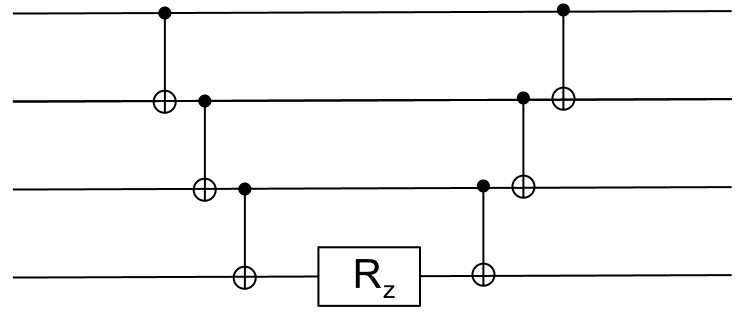}}
 \caption[Exponentiation of tensor products of Pauli-Z matrices]{A demonstration of how to exponentiate tensor products of Pauli matrices. First, the parity of the four qubits is computed with CNOT gates, and then a single-qubit phase rotation $R_z$ is applied. Then, we uncompute the parity with three further CNOT gates.\label{zzzz}}
\end{figure}

\noindent The terms of both the Bravyi-Kitaev Hamiltonian and the Jordan-Wigner Hamiltonian can be broken into two subsets, where the terms in each subset all mutually commute but the subsets do not commute with one another. These groups are as follows:
\begin{eqnarray}
\hat{H}_{BK, Z} =\ -&0.81261\ \I+0.171201\ \sigma_{0}^z\ +0.16862325\ \sigma_{1}^z\ -0.2227965\ \sigma_{2}^z\ +0.171201\ \sigma_{1}^z \sigma_{0}^z\  \nonumber \\
+&0.12054625\ \sigma_{2}^z\sigma_{0}^z\ +0.17434925\ \sigma_{3}^z\sigma_{1}^z\ +0.165868\ \sigma_{2}^z\sigma_{1}^z\sigma_{0}^z\ \nonumber \\
+&0.12054625\ \sigma_{3}^z\sigma_{2}^z\sigma_{0}^z\ -0.2227965\ \sigma_{3}^z\sigma_{2}^z\sigma_{1}^z\ +0.165868\ \sigma_{3}^z\sigma_{2}^z\sigma_{1}^z\sigma_{0}^z; \\
& \nonumber \\
\hat{H}_{BK, XY} = \ &0.04532175\ \sigma_{2}^x\sigma_{1}^z\sigma_{0}^x\ +0.04532175\ \sigma_{2}^y\sigma_{1}^z\sigma_{0}^y\ +0.04532175\ \sigma_{3}^z\sigma_{2}^x\sigma_{1}^z\sigma_{0}^x\ \nonumber \\
+&0.04532175\ \sigma_{3}^z\sigma_{2}^y\sigma_{1}^z\sigma_{0}^y;
\end{eqnarray}
\begin{eqnarray}
\hat{H}_{JW, Z} =\ -&0.81261\ \I+0.171201\ \sigma_{0}^z\ +0.171201\ \sigma_{1}^z\ -0.2227965\ \sigma_{2}^z\ -0.2227965\ \sigma_{3}^z\  \nonumber \\
+&0.16862325\ \sigma_{1}^z\sigma_{0}^z\ +0.12054625\ \sigma_{2}^z\sigma_{0}^z\ +0.165868\ \sigma_{2}^z\sigma_{1}^z\ +0.165868\ \sigma_{3}^z\sigma_{0}^z\ \nonumber \\
+&0.12054625\ \sigma_{3}^z\sigma_{1}^z\ +0.17434925\ \sigma_{3}^z\sigma_{2}^z; \\
& \nonumber \\
\hat{H}_{JW, XY} = -&0.04532175\ \sigma_{3}^x\sigma_{2}^x\sigma_{1}^y\sigma_{0}^y\ +0.04532175\ \sigma_{3}^x\sigma_{2}^y\sigma_{1}^y\sigma_{0}^x\ +0.04532175\ \sigma_{3}^y\sigma_{2}^x\sigma_{1}^x\sigma_{0}^y\ \nonumber \\
-&0.04532175\ \sigma_{3}^y\sigma_{2}^y\sigma_{1}^x\sigma_{0}^x.
\end{eqnarray}
To understand what computational resources are required for exponentiating operators of this kind, consider the example of the exponentiation of a fourfold product of $\sigma^z$ matrices, $e^{i(\sigma^z \otimes \sigma^z  \otimes \sigma^z \otimes \sigma^z)}$, which is depicted in a circuit diagram in Figure~\ref{zzzz}~\cite{MikeIke}.

In general, an $n$-fold tensor product of Pauli-Z matrices will require $2(n-1)$ CNOT gates and one single-qubit gate (SQG) to exponentiate on a quantum computer. If there are Pauli-X or -Y matrices in the tensor product, we must apply the single-qubit Hadamard or $R_x$ gate to change basis to the $X$ or $Y$ basis, respectively, before we compute the parity of the set of qubits with CNOT's, and also apply the inverse gates as part of the uncomputing stage \cite{MikeIke}. These gates are given by:
\begin{equation}
H = \frac{1}{\sqrt{2}} \left[\begin{array}{cc} 1&1\\1 & -1 \end{array}\right] \quad \quad \quad R_x = \frac{1}{\sqrt{2}} \left[ \begin{array}{ll} 1&i\\i & 1 \end{array}\right] 
\end{equation}
Thus, each non-$\sigma^z$ term in a tensor product of Pauli matrices adds $2$ single-qubit gates to the cost of exponentiation. For example, the circuit for exponentiating the term $\sigma_{3}^y\sigma_{2}^x\sigma_{1}^x\sigma_{0}^y$ is depicted in Figure~\ref{yxxy}

\begin{figure}[!h]
\centerline{ \includegraphics[scale = .35]{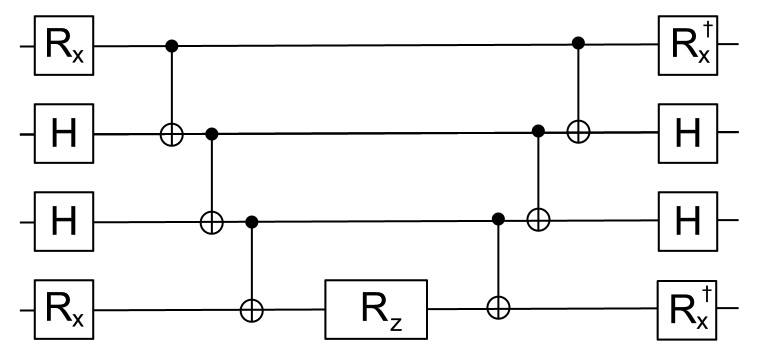}}
 \caption[\label{yxxy} Exponentiation of tensor products of Pauli-X and -Y matrices]{A demonstration of how to exponentiate tensor products of Pauli-X and -Y matrices. First, the qubits are put in the correct basis by the application of $R_x$ or Hadamard gates. Then, the parity of the four qubits is computed with CNOT gates, and then a single-qubit phase rotation $R_z$ is applied. Then, we uncompute the parity with more CNOT gates, and finally change back to the computational (Z) basis.}
 \label{yxxy}
\end{figure}
Using the resource counting methods detailed above, we can count the number of single-qubit gates (SQG's) and CNOT gates required to exponentiate (for arbitrary propagation time) the subsets of the Hamiltonians for both encodings. The results of this analysis are in Table~\ref{tab4}.

\begin{table}[h!]
\begin{center}
\begin{tabular}{| c |c| c | c|}
\hline
 & SQG's & CNOT's & Totals \\
\hline
$\hat{H}_{BK,Z}$ & 10 & 24 & 34 \\
\hline
$\hat{H}_{BK,XY}$ & 20 & 20 & 40 \\
\hline
Totals & 30 & 44 & {\bf 74}\\
\hline
\hline
$\hat{H}_{JW,Z}$ & 10 & 12 & 22 \\
\hline
$\hat{H}_{JW,XY}$ & 36 & 24 & 60 \\
\hline
Totals & 46 & 36 & {\bf 82} \\
\hline
\end{tabular}
\caption{\label{tab4}The number of single-qubit gates and CNOT gates required to exponentiate subsets of the electronic Hamiltonian for the hydrogen molecule, represented in terms of spin variables through either the Bravyi-Kitaev transformation or the Jordan-Wigner transformation.}
\end{center}
\end{table}

\begin{figure}[!h]
\centerline{ \includegraphics[scale = .6]{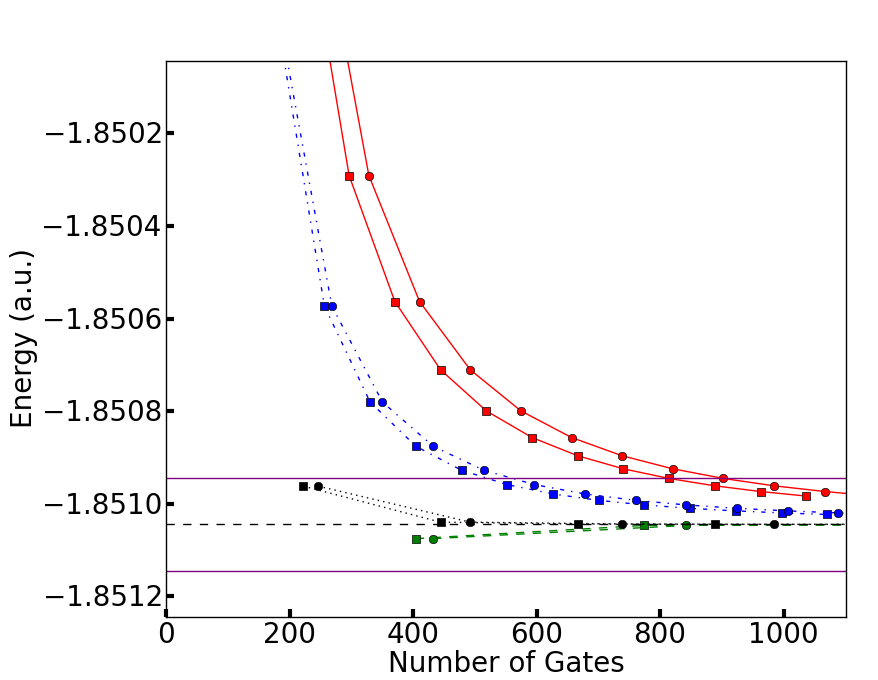}}
 \caption[Eigenvalue approximation]{\label{energy}The approximation to the ground state eigenvalue, for both the Bravyi-Kitaev Hamiltonian (squares) and Jordan-Wigner Hamiltonian (circles), as a function of the number of gates required. The solid curves are the first order Suzuki-Trotter approximations, the dot-dashed second order, the dotted third order, and the dashed fourth. The dotted horizontal line represents the true eigenvalue, while the solid lines above and below represent the bounds for chemical precision.}
\end{figure}

We now have the tools to compare the number of gates required to compute the ground state eigenvalue of either the Bravyi-Kitaev Hamiltonian or the Jordan-Wigner Hamiltonian to chemical precision ($\pm 10^{-4}$ a.u). Due to the small size of our model of the hydrogen system, it is easy for a classical computer to simulate the behavior of the quantum simulator. The true propagator $U = e^{-i \hat{H} t}$ can be computed to sufficient precision by a matrix exponential function in Mathematica or a similar software package. Time evolution of the ground state by the true propagator will result in phase evolution:
\begin{equation}
U \ket{\psi_g} = e^{-i E_g t} \ket{\psi_g}.
\end{equation}
We can therefore compute the exact eigenvalue as follows:
\begin{equation}
\bra{\psi_g} U \ket{\psi_g} = \bra{\psi_g} e^{-i E_g t} \ket{\psi_g} = e^{-i E_g t}.
\end{equation}
We set the propagation time to unity, and extract the true eigenvalue $E_g$ from the complex phase $e^{-i E_g}$. To approximate the eigenvalue, we use a Suzuki-Trotter approximation to the true propagator, $\tilde{U}$, and perform an analogous procedure:
\begin{equation}
\frac{\bra{\psi_g} \tilde{U} \ket{\psi_g}}{|\bra{\psi_g} \tilde{U} \ket{\psi_g}|} =  e^{-i \tilde{E}_g t}.
\end{equation}
The approximation to the true ground state eigenvalue, $\tilde{E}_g$, becomes better as we increase the number of Trotter steps $n$. Figure~\ref{energy} below plots the estimated eigenvalues of the minimal basis Jordan-Wigner and Bravyi-Kitaev Hamiltonians as a function of the number of gates required, for the first four orders of Suzuki-Trotter formulae.

We now compare this result to previous estimates. The benchmark is the gate count given in \cite{James} for approximating the Jordan-Wigner Hamiltonian's ground state eigenvalue. It is clear from Figure~\ref{energy} that our first order approximation requires $\approx900$ gates to obtain chemical precision for the Jordan-Wigner Hamiltonian, while the gate estimate in \cite{James} was about $500$ for the same task. This discrepancy arises from the fact that any number of variants on the first order Suzuki-Trotter formula could have been used in \cite{James}. Given a noncommuting set of Hamiltonian terms, there is some optimal ordering that will produce the best accuracy. It is not possible to know in advance which ordering is optimal, and given that the number of terms in an electronic Hamiltonian scales as $O(n^4)$, in general it is difficult to optimize over the space of possible orderings. We have used the most na\"{i}ve variant of the first order Suzuki-Trotter formula in Figure~\ref{energy}:
\begin{equation}
e^{-i \hat{H} t} = e^{-i(\hat{H}_Z + \hat{H}_{XY}) t} \approx (e^{-i \hat{H}_Z \frac{t}{n}} e^{-i \hat{H}_{XY} \frac{t}{n}})^n.
\end{equation}

\begin{figure}[!h]
\centerline{ \includegraphics[scale = .6]{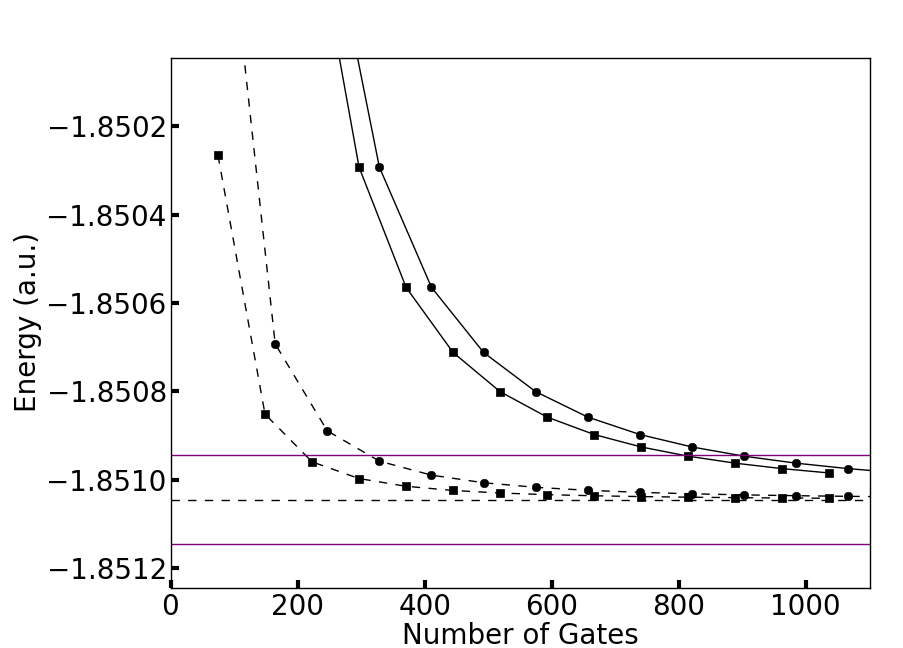}}
 \caption[Eigenvalue approximation]{\label{approx}The approximation to the ground state eigenvalue, for both the Bravyi-Kitaev Hamiltonian (squares) and Jordan-Wigner Hamiltonian (circles), as a function of the number of gates required. The solid curve is the na\"{i}ve first order Suzuki-Trotter approximation, while the dashed curve is the result from alternating the noncommuting terms. The dotted horizontal line represents the true eigenvalue, while the solid lines above and below represent the bounds for chemical precision. The ground state eigenvalue of the Bravyi-Kitaev Hamiltonian can be approximated to chemical precision with 222 gates, while it takes 328 gates to do the same for the Jordan-Wigner Hamiltonian.}
\end{figure}

However, due to the small size of our model of the hydrogen molecule, it is easy to find an ordering that produces better accuracy. A second, more sophisticated, variant of the first order formula is to arrange the terms in $\hat{H}_{Z}$ and $\hat{H}_{XY}$ in order of descending coefficient magnitude. For example, for the Bravyi-Kitaev Hamiltonian, we have:
\begin{eqnarray}
&\hat{H}_{Z}: \{h_{Z0}, h_{Z1}, h_{Z2}, \dots \} = \{ -0.81261\ \I , -0.2227965\ \sigma_{2}^z , -0.2227965\ \sigma_{3}^z\sigma_{2}^z\sigma_{1}^z , \dots \}; \\
&\hat{H}_{XY}: \{h_{XY0}, h_{XY1}, h_{XY2}, \dots \} = \{ 0.04532175\ \sigma_{2}^x\sigma_{1}^z\sigma_{0}^x, 0.04532175\ \sigma_{2}^y\sigma_{1}^z\sigma_{0}^y, \dots \}.
\end{eqnarray}
Then, we approximate the propagator by alternately exponentiating one term from the ordered list of $\hat{H}_{Z}$ terms and one term from the ordered list of $\hat{H}_{XY}$ terms until we have used all terms from $\hat{H}_{XY}$. Then we exponentiate the rest of $\hat{H}_{Z}$:
\begin{equation}
e^{-i \hat{H} t} \approx (e^{-i h_{Z0} \frac{t}{n}} e^{-i h_{XY0} \frac{t}{n}} e^{-i h_{Z1} \frac{t}{n}} e^{-i h_{XY1} \frac{t}{n}} \cdots e^{-i h_{XY3} \frac{t}{n}} e^{-i h_{Z4} \frac{t}{n}} e^{-i h_{Z5} \frac{t}{n}} \cdots)^n.
\end{equation}
With this method, we find that the number of gates required to obtain a chemical precision estimate of the ground state eigenvalue of the Jordan-Wigner Hamiltonian is $\approx 300$, fewer than the result from \cite{James}. Figure~\ref{approx} compares the eigenvalue approximations for the na\"{i}ve first order method and the more sophisticated variant.

\begin{figure}[!h]
\centerline{ \includegraphics[scale = .6]{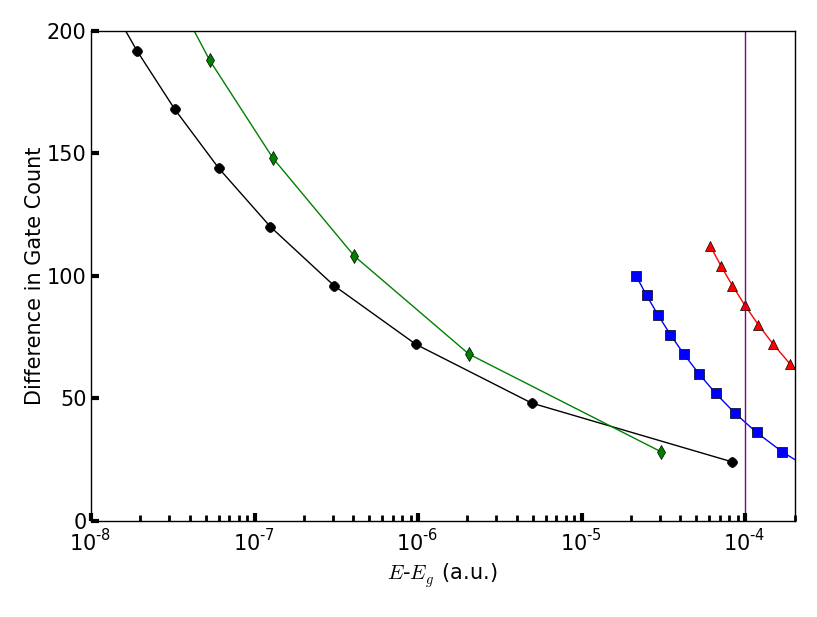}}
 \caption[Eigenvalue approximation]{\label{savings}The gate savings of using the Bravyi-Kitaev method instead of the Jordan-Wigner method, as a function of the precision in the estimate of the ground state eigenvalue for the first four orders of Suzuki-Trotter formulae. The vertical line is the threshold error for chemical precision. The triangle data points are first order, the squares second, the circles third, and the diamonds fourth.}
\end{figure}

The point is that the systematic advantage of the Bravyi-Kitaev method over the Jordan-Wigner method is not obscured by the kind of term-ordering optimization that we have demonstrated above. Exponentiating the Bravyi-Kitaev Hamiltonian requires $74$ gates per first order Trotter step (of any variant), while the Jordan-Wigner Hamiltonian requires $82$ gates per first order Trotter step. To obtain a precision of $\pm 10^{-4}$ a.u to the true eigenvalue with the na\"{i}ve first order Suzuki-Trotter approximation requires $11$ Trotter steps for both the Bravyi-Kitaev and Jordan-Wigner Hamiltonian, for a total cost of $814$ gates versus $902$ gates. With the noncommuting terms intermixed, it takes only $3$ Trotter steps to obtain the same precision for the Bravyi-Kitaev Hamiltonian, and 4 Trotter steps for the Jordan-Wigner Hamiltonian. Thus, if we intermix the noncommuting terms, the Bravyi-Kitaev transformation allows one to utilize $222$ gates instead of the $328$ gates required by the Jordan-Wigner transformation to obtain an equally precise estimate of the hydrogen molecule's ground state eigenvalue when using a first order Suzuki-Trotter approximation. When using higher-order Suzuki-Trotter approximations to obtain better than chemical precision, the gate savings increases (Fig.~\ref{savings}).

\section{Conclusions}\label{conc}

In this paper we have worked out a detailed application of the Bravyi-Kitaev transformation to Hermitian second quantized operators that appear in quantum chemical Hamiltonians. We suggest that this transformation should replace the Jordan-Wigner transformation for fermionic quantum simulation algorithms. We have demonstrated that the Bravyi-Kitaev transformation results in a small reduction in the number of gates, from $328$ gates to $222$ gates, required to implement a quantum simulation algorithm for electron dynamics in the simplest possible molecular system of H$_2$ in a minimal basis.

In some sense, molecular hydrogen in a minimal basis is a poor showcase of the power of the Bravyi-Kitaev transformation. Our description of this molecule utilizes four molecular orbitals, and hence four qubits. The spin Hamiltonians we derive using either the Bravyi-Kitaev transformation or the Jordan-Wigner Hamiltonian involve four-local Pauli tensor products, the result being that the cost of simulating time evolution under the Bravyi-Kitaev Hamiltonian on a quantum computer is only slightly reduced from that for the Jordan-Wigner Hamiltonian. However, were we to use a more sophisticated description of the H$_2$ --- for example, with eight molecular orbitals --- the Jordan-Wigner spin Hamiltonian would contain up to eight-local Pauli tensor products, while the Bravyi-Kitaev spin Hamiltonian would not. Given the asymptotically better $O(\log n)$ scaling of the Bravyi-Kitaev method  as compared to the $O(n)$ scaling of the Jordan-Wigner transformation, the difference between the two methods will become greater for larger basis sets and larger molecules --- the simulation of which is, after all, is the true goal of quantum simulation for quantum chemistry, since the small molecules are within the reach of conventional computers. However, by showing that the Bravyi-Kitaev method is more efficient for the smallest conceivable chemical system, we have demonstrated that there is no algorithmic overhead inherent to the Bravyi-Kitaev method that must be overcome by scaling up the size of problems to which it is applied. We have demonstrated the superior efficiency of the Bravyi-Kitaev transformation for all quantum chemical simulations. Thus, making use of the Bravyi-Kitaev transformation for fermionic quantum simulation will make simulations of larger molecules and with larger basis sets more readily accessible to experiment.

 \section{Acknowledgments}
The authors thank the Aspuru-Guzik group for their hospitality during the summers of 2011 and 2012, when parts of this work were completed. We are indebted to Jarod Maclean, John Parkhill, Sam Rodriques, Joshua Schrier, Robert Seeley, and James Whitfield for productive discussions. This project is supported by NSF CCI center, ``Quantum Information for Quantum Chemistry (QIQC)", award number CHE-1037992, by NSF award PHY-0955518 and by AFOSR award no FA9550-12-1-0046.

\end{document}